\documentclass[runningheads]{llncs}
\usepackage{graphicx}
\usepackage{paralist}
\usepackage{CJK}
\usepackage{listings}
\usepackage{xcolor}
\usepackage{array}
\usepackage{pgf,tikz}
\usetikzlibrary{positioning, shapes.geometric, shapes.symbols, shapes.arrows, arrows.meta, decorations.pathreplacing, calc, fit, backgrounds, matrix}
\usepackage{float}
\usepackage{enumitem}

\usepackage{amsthm}
\usepackage{amssymb}
\usepackage{algorithm}
\usepackage[noend]{algpseudocode}
\usepackage{xspace}

\usepackage{listings}
\usepackage{color}
\usepackage{mathtools}
\usepackage{multirow}
\usepackage{graphics}
\usepackage{paracol}
\usepackage{stmaryrd}
\usepackage{subcaption}
\usepackage{booktabs}
\usepackage{tabularx}
\usepackage{caption}
\usepackage[normalem]{ulem}
\usepackage{xurl}

\usepackage{hyperref}
\usepackage[capitalise]{cleveref}

\definecolor{dkgreen}{rgb}{0,0.6,0}
\definecolor{gray}{rgb}{0.5,0.5,0.5}
\definecolor{mauve}{rgb}{0.58,0,0.82}
\definecolor{brightcerulean}{rgb}{0.11,0.67,0.84}
\definecolor{cerulean}{rgb}{0.0, 0.48, 0.65}

\lstdefinestyle{mystyle}{frame=none,
  aboveskip=3mm,
  belowskip=3mm,
  showstringspaces=false,
  columns=flexible,
  basicstyle={\ttfamily\scriptsize},
  numbers=none,
  numberstyle=\tiny\color{gray},
  keywordstyle=\color{blue},
  commentstyle=\color{dkgreen},
  stringstyle=\color{mauve},
  breaklines=true,
  numbers=left,
  numbersep=5pt,
  upquote=true,
  breakatwhitespace=true,
  tabsize=2,
  captionpos=b,
  keywordstyle=[2]{\color{cerulean}},
  morekeywords=[2]{assume, assert, int_nd, bool_nd, is_deref, alloc, gep, store, load, memcpy, havoc},
  keywordstyle=[3]{\color{orange}},
  morekeywords=[3]{@len, @char, @cap, @buf, @}
}

\newcommand{\ag}[1]{\textcolor{purple}{\textbf{Arie:} #1}}
\newcommand{\jorge}[1]{\textcolor{orange}{\textbf{Jorge:} #1}}
\newcommand{\yusen}[1]{\textcolor{blue}{\textbf{Yusen:} #1}}
\newcommand{\ig}[1]{\textcolor{brown}{\textbf{Isabel:} #1}}
\newcommand{\todo}[1]{\textcolor{red}{\textbf{TODO:} #1}}

\renewcommand{\ag}[1]{}
\renewcommand{\jorge}[1]{}
\renewcommand{\yusen}[1]{}
\renewcommand{\ig}[1]{}
\renewcommand{\todo}[1]{}

\newcommand{\m}[1]{\mathsf{#1}}
\newcommand{\mi}[1]{\mathit{#1}}
\newcommand{\mt}[1]{\mathtt{#1}}
\newcommand{\Zrepeat}[2]{\foreach \n in {1,...,#1}{#2}}
\newcommand{\ind}[1][1]{\Zrepeat{#1}{\quad}}

\newcommand{\tdesignation}{\mathsf{\sharp}}

\newcommand{\tool}[1]{\textsc{#1}\xspace}
\newcommand{\lang}[1]{\texttt{#1}\xspace}
\newcommand{\crab}{\tool{Crab}}
\newcommand{\seabmc}{\tool{SeaBMC}}
\newcommand{\seadsa}{\tool{SeaDsa}}
\newcommand{\crabir}{\lang{CrabIR}}
\newcommand{\memModel}{RUMM\xspace}
\newcommand{\zthree}{\tool{Z3}}
\newcommand{\ytwo}{\tool{Yices2}}
\newcommand{\stacklabel}[1]{\stackrel{\smash{\scriptscriptstyle \mathrm{#1}}}}
\newcommand{\Def}{\stacklabel{\mathrm{def}}}

\newcommand{\script}[1]{\mathcal{#1}}
\newcommand{\scriptV}{\script{V}}

\newcommand{\pair}[2]{#1 \times #2}

\newcommand{\prog}{\m{P}}
\newcommand{\func}{\m{F}}
\newcommand{\basicblock}{\m{BB}}
\newcommand{\statement}{\m{S}}

\newcommand{\rgnstatement}{\m{\statement_{ptr}}}

\newcommand{\expression}{\m{E}}
\newcommand{\intexpression}{\m{\expression_{int}}}

\newcommand{\conditionexpression}{\m{\expression_{cond}}}

\newcommand{\constant}{\m{Const}}
\newcommand{\op}{\m{op}}
\newcommand{\opint}{\op_{int}}

\newcommand{\opcmp}{\op_{cmp}}

\newcommand{\kdeclare}{\m{declare}}

\newcommand{\kreturn}{\lang{return}}
\newcommand{\kgoto}{\lang{goto}}
\newcommand{\funcname}{\mathit{fun}}
\newcommand{\true}{\m{true}}
\newcommand{\false}{\m{false}}
\newcommand{\loadref}{\m{load}}
\newcommand{\makeref}{\m{alloc}}
\newcommand{\storeref}{\m{store}}

\newcommand{\gepref}{\m{gep}}

\newcommand{\bblabel}{l}

\newcommand{\cassume}{\lang{assume}}
\newcommand{\cassert}{\lang{assert}}

\newcommand{\stat}{\sigma}

\newcommand{\absstat}{\sigma^{\tdesignation}}
\newcommand{\cstat}{\m{State}}

\newcommand{\astat}{\cstat^{\tdesignation}}

\newcommand{\baseaddr}{\mi{baddr}}

\newcommand{\offset}{\mi{offset}}
\newcommand{\sz}{\mi{sz}}
\newcommand{\mb}{\mi {mb}}

\newcommand{\cell}{\mi {cell}}
\newcommand{\Cell}{\m{Cell}}

\newcommand{\mem}{\mi {mem}}

\newcommand{\cacheused}{\mi {used}}
\newcommand{\cachedirty}{\mi {dirty}}
\newcommand{\cacheinitpack}{\mi {ispk}}
\newcommand{\cachebase}{\mi{{cache}^{base}}}

\newcommand{\varint}{\m{num}}
\newcommand{\varscalar}{\m{scl}}

\newcommand{\varptr}{\m{ptr}}
\newcommand{\varptrone}{\varptr\m{1}}
\newcommand{\varptrtwo}{\varptr\m{2}}
\newcommand{\varptrbase}{\varptr^{\mathit{base}}}
\newcommand{\varptronebase}{\varptrone^{\mathit{base}}}
\newcommand{\varptrtwobase}{\varptrtwo^{\mathit{base}}}

\newcommand{\varfld}{\m{fld}}
\newcommand{\varfldone}{\varfld\m{1}}
\newcommand{\varfldtwo}{\varfld\m{2}}
\newcommand{\vfld}{\scriptV_{\mathit{fld}}}

\newcommand{\objenv}{\mi {flag}}
\newcommand{\ObjEnv}{\m{Flag}}
\newcommand{\AbsObjEnv}{\ObjEnv^{\tdesignation}}
\newcommand{\absobjenv}{\objenv}

\newcommand{\cache}{\mi {cache}}
\newcommand{\Cache}{\m{Cache}}
\newcommand{\abscache}{\cache}
\newcommand{\AbsCache}{\Cache^{\tdesignation}}
\newcommand{\Summary}{\m{Summary}}
\newcommand{\summary}{\mi {sum}}
\newcommand{\Storage}{\m{Storage}}
\newcommand{\storage}{\mi {storage}}
\newcommand{\abssummary}{\summary}
\newcommand{\AbsSummary}{\Summary^{\tdesignation}}

\newcommand{\vint}{\scriptV_{\mathit{int}}}
\newcommand{\vscalar}{\scriptV_{\mathit{scl}}}

\newcommand{\fldenv}{\mi {fields}}
\newcommand{\FldEnv}{\m{FldVal}}

\newcommand{\vptr}{\scriptV_{\mathit{ptr}}}

\newcommand{\vptrbase}{\vptr^{\mathit{base}}}

\newcommand{\ScalarEnv}{\m{Scalar}}
\newcommand{\scalarenv}{\mi{scalar}}

\newcommand{\tran}[2]{\llbracket#1\rrbracket^{\scriptscriptstyle #2}}
\renewcommand{\gets}{\mathbin{=}}
\newcommand{\cassign}[2]{\Tlet #1 \gets #2 \Tin}

\newcommand{\cassigntwo}[4]{\Tlet #1 \gets #2 \Tand #3 \gets #4 \Tin}
\newcommand{\kw}[1]{\boldsymbol{\mathsf{#1}}} 

\newcommand{\Tand}{\;\kw{and}\;}

\newcommand{\Tlet}{\kw{let}\;}
\newcommand{\Tin}{\;\kw{in}\;}

\newcommand{\Tval}{\mi{val}}

\newcommand{\allocator}{\mt{allocator}}

\newcommand{\addEquality}{\mt{addEqual}}
\newcommand{\testEquals}{\mt{equals}}

\newcommand{\invalidateCacheIfMiss}{\mt{cacheSync}}

\newcommand{\Basedom}{{\m{Scalar}}^\sharp}

\newcommand{\Bool}{\m{Bool}}

\newcommand{\booldom}{{{\Bool}^\sharp}}

\newcommand{\eqdommacro}[1]{\m{E^{\sharp}_{#1}}}

\newcommand{\eqdom}{\m{Eq}}
\newcommand{\numdom}{\m{Num}}
\newcommand{\AddrsDom}{\eqdommacro{p}}
\newcommand{\eqregsdom}{\eqdommacro{s}}
\newcommand{\eqfldsdom}{\eqdommacro{f}}
\newcommand{\eqsfdom}{\eqdommacro{sf}}

\newcommand{\regseq}{\mi {e_s}}
\newcommand{\fldseq}{\mi {e_f}}
\newcommand{\regfldseq}{\mi {e_{sf}}}
\newcommand{\addrseq}[1][]{\mi{e_p}}

\newcommand{\MemoryBank}{\m{MB}}
\newcommand{\Memory}{\m{Memory}}
\newcommand{\AbsMemoryBank}{\MemoryBank^\sharp}
\newcommand{\AbsMemory}{\Memory^\sharp}
\newcommand{\memory}{\mi{mem}}
\newcommand{\absmemory}{\memory}

\newcommand{\mlocate}{\ensuremath{\m{findmb}}\xspace}
\newcommand{\mlocateabs}{\ensuremath{\mlocate^\sharp}\xspace}

\newcommand{\code}[1]{\lstinline[basicstyle=\footnotesize\ttfamily] {#1}\xspace}
\newcommand{\hex}[1]{\ensuremath{#1_{16}}}

\theoremstyle{definition}
\newtheorem{examplex}{Example}[subsubsection]
\renewcommand{\theexamplex}{%
  \ifnum\value{subsubsection}>0 \thesubsubsection \else
  \ifnum\value{subsection}>0 \thesubsection \else
  \thesection \fi\fi .\arabic{examplex}%
}

\newcommand{\mrudomain}{MRUD\xspace}
\newcommand{\neweqdomain}{VarEq domain\xspace}

\renewcommand{\themydef}{%
  \ifnum\value{subsubsection}>0 \thesubsubsection \else
  \ifnum\value{subsection}>0 \thesubsection \else
  \thesection \fi\fi .\arabic{mydef}%
}

\renewcommand{\themytheorem}{%
  \ifnum\value{subsubsection}>0 \thesubsubsection \else
  \ifnum\value{subsection}>0 \thesubsection \else
  \thesection \fi\fi .\arabic{mytheorem}%
}

\algdef{SE}[FUNCTION]{MyFunction}{EndMyFunction}[2]{#1(#2)~$\equiv$}{\textbf{end}}
\algtext*{EndMyFunction}

\algdef{SE}[LET]{Let}{EndLet}[2]{$\Tlet$\ #1\ $\gets$\ #2\ }{$\Tin$}

\crefname{cline}{line}{lines}
\crefname{irline}{line}{lines}

\begin{document}
\title{Automatic Inference of Relational Object Invariants}
%
%
\author{Yusen Su\inst{1}\orcidID{0009-0004-8813-0797} \and
Jorge A. Navas\inst{2}\orcidID{0000-0002-0516-1167} \and
Arie Gurfinkel\inst{1}\orcidID{0000-0002-5964-6792} \and Isabel Garcia-Contreras\inst{1,3}\orcidID{0000-0001-6098-3895}}
\authorrunning{Y. Su et al.}
%
\institute{Department of Electrical and Computer Engineering, University of Waterloo \and Certora Inc. \and{Black Duck Software, Inc.}}
\maketitle              
\begin{abstract}
Relational object invariants (or representation invariants) are relational
properties held by the fields of a (memory) object throughout its lifetime.
For example, the length of a buffer never exceeds its capacity. Automatic
inference of these invariants is particularly challenging because they are
often broken temporarily during field updates.

In this paper, we present an Abstract Interpretation-based solution to infer
object invariants. Our key insight is a new object abstraction for memory
objects, where memory is divided into multiple \emph{memory banks}, each
containing several objects. Within each bank, objects are abstracted by
separating the \emph{most recently used} (MRU) object, represented precisely
with strong updates, while the rest are summarized. For an effective
implementation of this approach, we introduce a new composite abstract domain,
which forms a reduced product of numerical and equality sub-domains. This
design efficiently expresses relationships between a small number of variables
(e.g., fields of the same abstract object).

We implement the new domain in the \crab abstract interpreter and evaluate it
on several benchmarks for memory safety. We show that our approach is
significantly more scalable for relational properties than the existing
implementation of \crab. To evaluate precision, we have integrated our analysis
as a pre-processing step to \seabmc bounded model checker, and show that it is
effective at both discharging assertions during pre-processing, and
significantly improving the run-time of \seabmc.
 
\end{abstract}
\section{Introduction}

Program invariants are crucial to capture properties that persist during runtime. Verifying programs with classes or data structures requires determining \emph{representation invariants} \cite{meyer1997}  that express \emph{consistency} properties (e.g., the length of a vector never exceeds its capacity) of those data types. For memory objects, representation invariants as \emph{object invariants} describe relational properties among object fields that hold across all program states where these objects are alive. These invariants are essential for proving memory safety and functional correctness of a program. 
However, the invariants become imprecise when the static analyzer is uncertain about which memory objects are affected by field updates, typically represented as \emph{weak} updates.

\begin{figure}[t]
\begin{minipage}[c]{0.5\linewidth}
\centering
\lstset{escapeinside={(*@}{@*)}}
\begin{lstlisting}[basicstyle=\footnotesize, language=C, style=mystyle]
#define N 100
struct byte_buf {
    int len;
    int cap;
    char *buf;
};
int main() {
    struct byte_buf *ary[N];(*@\label[cline]{cline:array}@*)
    for (int i = 0; i < N; ++i) {
        struct byte_buf *p = malloc(sizeof(struct byte_buf)); (*@\label[cline]{cline:start}@*)
        int sz = i + 1;
        p->len = i; p->cap = sz; (*@\label[cline]{cline:others}@*)
        p->buf = malloc(sz); (*@\label[cline]{cline:buffer}@*)
        ary[i] = p; (*@\label[cline]{cline:state}@*)
    }
    char *new_buf = malloc(20);
    ary[0]->len = 15;(*@\label[cline]{cline:length}@*)
    ary[0]->cap = 20;(*@\label[cline]{cline:capacity}@*)
    ary[0]->buf = new_buf;(*@\label[cline]{cline:ubuffer}@*)
    assert(ary[0]->len <= ary[0]->cap);(*@\label[cline]{cline:assert}@*)
    ary[0]->buf[ary[0]->len] = '\0';(*@\label[cline]{cline:mem}@*)
}
\end{lstlisting}
\caption{A simple C program.}
\label{fig:discuss-example}
\end{minipage}
\begin{minipage}[c]{0.44\linewidth}
\centering
  \includegraphics[scale=0.71]{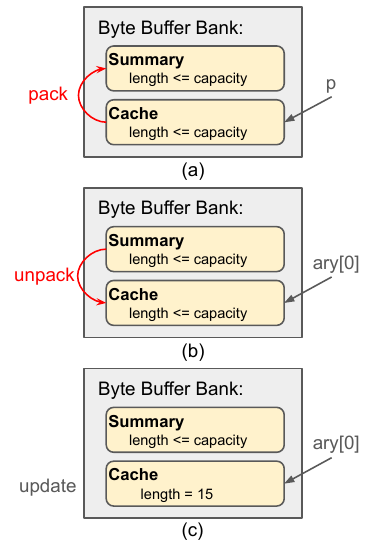}
  \caption{Abstract memory state on \cref{cline:length} of \cref{fig:discuss-example}.}
  \label{fig:cache}
\end{minipage}
\vspace{-1em}
\end{figure}

Consider a C program in  \cref{fig:discuss-example} that uses a \code{byte\_buf} to represent a resizable byte buffer with length and capacity. The program keeps an array \code{ary} of byte buffers. Each initialized element of \code{ary} satisfies an invariant: \code{len <= cap}. Discovering this invariant is crucial for establishing memory safety (e.g., proving safe access on~\cref{cline:mem}), yet, notoriously hard for abstract interpreters. Note that \emph{recency}~\cite{DBLP:conf/sas/BalakrishnanR06} does not help here because all memory stores after the \code{for} loop are modeled as weak updates. For instance, Mopsa~\cite{DBLP:conf/tacas/MonatOM23} with recency does not prove the assertion on~\cref{cline:assert}, since the inferred invariant is $len > 0 \land cap > 1$.

In this paper, we present a new technique for inferring object invariants.
We capture field updates \emph{strongly} in a separate temporary object abstraction and join it with previously established invariants only when necessary. While preserving soundness, our approach produces more precise analysis results by not weakening inferred invariants with intermediate object states between updates.

First, we introduce a new concrete memory model that organizes memory as a collection of \emph{memory banks}, each containing certain memory objects. The partitioning is achieved by a parameterized function that assigns each memory object in the program a corresponding bank. Each bank has two components: \emph{storage}, holding objects, and \emph{cache}, storing the object being read from or written to. For example, all byte buffers in \cref{fig:discuss-example} are placed into the storage of the same bank. The field updates on \cref{cline:others} require loading the byte buffer referred by pointer \code{p} into the cache before updates. The cache singles out the object being modified. For brevity, we specify this usage pattern with a size of one as \emph{most recently used} (MRU) and denote the object in the cache as the MRU object.

Second, we follow a standard summarization-based abstraction with a single summary object with its invariants representing properties common to all the objects stored in each bank. Similar to the concrete model, all memory updates are handled through the MRU object. This avoids temporarily breaking the invariants of the (abstract) summary object, as changes to the MRU object do not impact the summarized invariants until it is merged back. \cref{fig:cache} presents the changes in the abstract memory state at \cref{cline:length}. The memory bank for byte buffers includes one MRU object and one summary object. Before evaluating \cref{cline:length}, as shown in \cref{fig:cache}(a), \code{p} refers to the MRU object, since the last two field updates (\cref{cline:others}) happened on this object. Following the initialization loop, \code{len <= cap} is kept for both MRU and summary objects.

The cache may \emph{miss} if the cached object is no longer the MRU. For example, the field update, \code{ary[0]->len = 15}, on \cref{cline:length} requires access to the byte buffer referenced by \code{ary[0]}, while the cache still holds the object referred by \code{p}. In this case, the cached object is \emph{packed} back to the summary (see \cref{fig:cache}(a)) and the new MRU object is \emph{unpacked} from the summary (\cref{fig:cache}(b)). We track pointer alias information to decide when to pack and unpack. Before each memory access, if the dereferenced pointer does not alias with the pointer accessed to the MRU object, packing and unpacking occur. In this example, after the loop computation, \code{p} does not alias \code{ary[0]}.

After the cache is replaced, the field update, \code{ary[0]->len = 15}, breaks the invariant \code{len <= cap}, but our solution (\cref{fig:cache}(c)) ensures that we update the content of the MRU object properly without affecting the invariants in the summary object. Then, the invariant is restored at~\cref{cline:capacity}, thus proving the assertion on~\cref{cline:assert} and memory safety on~\cref{cline:mem} through our invariants in the cache.

Third, we introduce a new abstract domain, called \emph{\mrudomain}, that infers automatically object invariants based on our new memory model. This domain requires combining heap (memory abstraction), must alias (flow-sensitive points-to information) and value (numerical relational invariants) analyses.  
Using a monolithic numerical domain is highly inefficient because of the large number of dimensions required to model  all program variables and their ghost versions that keep track of base addresses, offsets, etc. However, a key insight is that each transfer function typically affects a small subset of variables (e.g.,  reading a field only updates the corresponding integer/pointer value).
Based on this observation, \emph{\mrudomain} is designed as a composite abstract domain where each memory bank is modeled separately and the propagation of facts between them is carefully limited to a small set of shared variables. This modular design is what makes \mrudomain both scalable for large code bases and capable of preserving precise object invariants.

We implemented \mrudomain in the \crab analyzer~\cite{DBLP:conf/vstte/GurfinkelN21} and evaluated both its scalability and precision. For scalability, we compare it to the summarization-based abstract domain implemented in \crab. Our approach shows improved scalability, with $75$X faster performance than the state-of-the-art. For precision, we compare it to the recency domain implemented on Mopsa using a small set of benchmarks. The results show that our approach successfully proves all assertions in the programs and achieves better precision by preserving object invariants. Additionally, we use \mrudomain in a case study with the bounded model checker \seabmc, where it effectively proves and discharges memory safety checks to reduce the verification cost of \seabmc.

In summary, the contributions of this paper are: 
\begin{inparaenum}[(1)]
\item We introduce a new memory model designed for object abstraction as an alternative to the C memory model, and describe the concrete semantics of an intermediate representation based on the new model (\cref{sec:mem-model-concrete});
\item We describe the \mrudomain and corresponding abstract transfer functions, and introduce a domain reduction for invariant refinement (\cref{sec:abs-sem});
\item We detail our implementation (\cref{sec:implementation}) and evaluate it in the \crab analyzer (\cref{sec:experiments}).
\end{inparaenum}
\section{Preliminaries}

\newcommand{\cV}{\mathcal{V}}
\newcommand{\cP}{\mathcal{P}}

Without loss of generality, we assume that the input program is in
\crabir~\cite{DBLP:conf/vstte/GurfinkelN21} intermediate representation. The
syntax of \crabir is shown in \cref{fig:lang-syntax}. We assume that each memory
object is a collection of integer and pointer fields. A pointer
is a pair of a base address and an offset, where an offset is given by a number $\varint$ and an optional field name $\varfld$. All named fields have fixed offsets. That is, field names are redundant -- they are use to simplify the abstraction function in the abstract semantics. In our implementation, the field names are automatically discovered by a whole-program pointer analysis during compilation from the source language to \crabir.

We write $\cV$ for the set of all program variables. The set $\cV$ is
partitioned into: integers $\vint$, pointers $\vptr$, and fields $\vfld$. The
union of $\vint$ and $\vptr$ is called \emph{scalars}. The statements in \crabir
consist of gotos, assumptions, assertions, and arithmetic and memory operations. All
statements are strongly typed. 
Allocation of memory objects is performed by
$\makeref$ (allocate). Pointer
arithmetic is handled by the $\gepref$ instruction that computes a destination
address using the base pointer and an integer offset. Memory reads and writes are done by $\loadref$ and
$\storeref$, respectively. As usual, a program $\cP$ is a control flow graph
(CFG) whose basic blocks are annotated with statements
from~\cref{fig:lang-syntax}.  \crabir also supports C-like memory
objects and it does not require them to be partitioned into fields. These are handled as in prior work~\cite{DBLP:conf/vstte/GurfinkelN21}. We omit such objects in the theoretical exposition in the paper, but handle them as in~\cite{DBLP:conf/vstte/GurfinkelN21} in our implementation.

\begin{figure}[t]
\centering
\scalebox{0.9}{
\begin{minipage}{\textwidth}
\begin{alignat*}{3}
\prog &::= \; \func^+ & \rgnstatement &::= \; \varptr := \makeref(\varfld, \varint) \mid \\
\func &::= \; \kdeclare \; \funcname (v^*) \{\; \basicblock^+ \;\} &&\ind[2] \varptrtwo, \varfldtwo := \gepref(\varptrone, \varfldone, \varint) \mid \\
\basicblock &::= \bblabel: \statement^* \;\kgoto \;\bblabel^+ \mid && \ind[2] \varscalar := \loadref(\varptr, \varfld) \mid \storeref(\varptr, \varfld, \varscalar) \\
&\ind[2] \bblabel: \statement^* \;\kreturn\; v^* & \intexpression &::= \; \constant \mid \varint \mid \intexpression \; \opint \; \intexpression \\
\statement &::= \;  \cassert(\conditionexpression) \mid \cassume(\conditionexpression) \mid \ind[3] & \;\conditionexpression &::= \intexpression \; \opcmp \; \intexpression \\
&\ind[2] \varint := \intexpression \mid \rgnstatement &&  %
\end{alignat*}
\end{minipage}
}
\caption{The syntax of \crabir.}
\label{fig:lang-syntax}
\vspace{-1em}
\end{figure}
We assume the reader is familiar with a standard numerical abstract domain that
provides the following operations: join ($\sqcup$), meet ($\sqcap$), widen
($\triangledown$), projection ($\m{project}(d, \scriptV)$) that projects an
abstract value $d$ to the variable set $\scriptV$, forget ($\m{forget}(d, v)$)
that removes a variable $v$ from an abstract value $d$, and constrain
($\m{addCons}(d, c)$) that restricts an abstract value $d$ by a linear
constraint $c$.

We use an equality domain over variable sets $\cV$ to express equivalence
relations such as $x \approx y$. The equality domain can be implemented using
weakly relational numerical domains (e.g.,
\cite{DBLP:journals/acta/Karr76,DBLP:conf/pado/Mine01,DBLP:conf/wcre/Mine01}).
We assume the equality domain has the following special operations:
$\addEquality$ for adding an equality, $\testEquals$ for testing whether an
abstract value entails an equality, and $\m{toCons}$ for
computing closed form of all implied equalities.

\section{Recent-Use Memory Model}
\label{sec:mem-model-concrete}

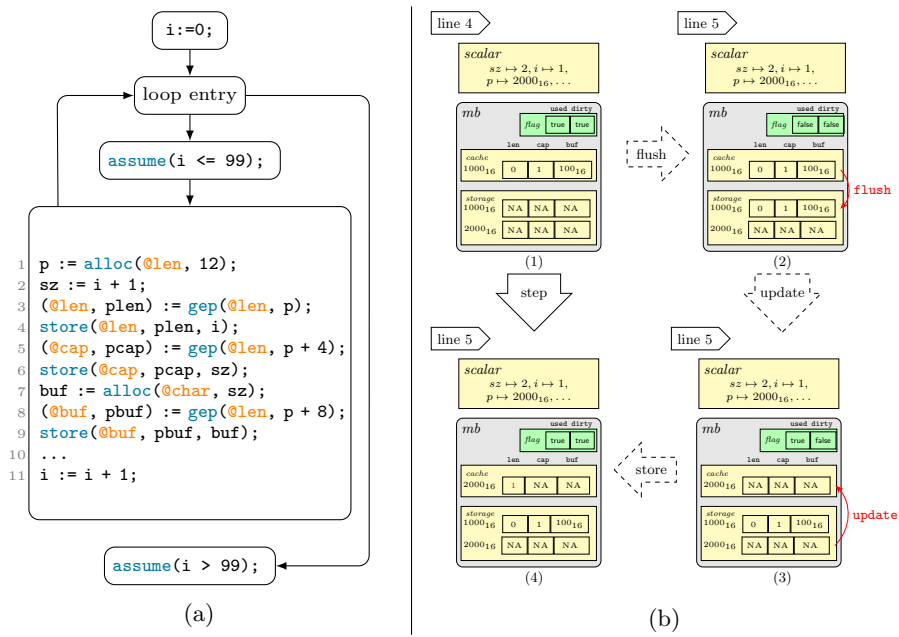
\begin{figure}[t]
\begin{minipage}{.45\textwidth}
\begin{subfigure}{\textwidth}\centering%
\lstset{escapeinside={(*@}{@*)}}
\begin{tikzpicture}[auto,
  box/.style={rectangle, rounded corners, draw, fill=white, align=center, minimum width=1cm, minimum height=0.5cm, font=\scriptsize},
  bodybox/.style={rectangle, rounded corners, draw, fill=white, align=center, font=\scriptsize}
  ]
 \node[box] (entry)    {\lstinline[basicstyle=\footnotesize, language=C, style=mystyle]
 {i:=0;}};
 \node[box, below= 0.35cm of entry] (cond)     {loop entry};
 \node[box, below= 0.35cm of cond] (header)     {
\lstinline[basicstyle=\footnotesize, language=C, style=mystyle]
 {assume(i <= 99);}
 };
 \node[bodybox, below=0.35cm of header, align=center, inner sep=4pt, text height=2.1cm, text width=4cm, minimum height=2.1cm, minimum width=4cm] (body)    {
\begin{minipage}{4.2cm}
\begin{lstlisting}[basicstyle=\footnotesize, language=C, style=mystyle]
p := alloc(@len, 12); (*@\label[irline]{irline:alloc}@*)
sz := i + 1;
(@len, plen) := gep(@len, p);
store(@len, plen, i); (*@\label[irline]{irline:store}@*)
(@cap, pcap) := gep(@len, p + 4); (*@\label[irline]{irline:gep}@*)
store(@cap, pcap, sz); 
buf := alloc(@char, sz); (*@\label[irline]{irline:absstate}@*)
(@buf, pbuf) := gep(@len, p + 8);
store(@buf, pbuf, buf); (*@\label[irline]{irline:oldstore}@*)
...
i := i + 1; (*@\label[irline]{irline:final}@*)
\end{lstlisting}
\end{minipage}
 };
 \node[box, below=0.35cm of body] (exit)    {
\lstinline[basicstyle=\footnotesize, language=C, style=mystyle]
 {assume(i > 99);}
 };
 \begin{scope}[rounded corners,-latex]
  \draw (cond.0) -- ([xshift=16.4mm]cond.east) --
  ([xshift=12.4mm]exit.east) -- (exit.0);
  \path (entry) edge (cond) (cond) edge (header) (header) edge (body);
  \draw (body.130) -- ([xshift=-10.2mm]cond.west) -- (cond); 
 \end{scope}
\end{tikzpicture}
\subcaption{}
\label{subfig:cfg}
\end{subfigure}
\end{minipage}
\vline\hfill
\begin{minipage}{.55\textwidth}
\begin{subfigure}{\textwidth}\centering%
\scalebox{0.6}{
\begin{tikzpicture}[
    basedomain/.style={rectangle, draw, fill=yellow!30, align=center, minimum width=1cm, minimum height=0.5cm, font=\scriptsize},
    composedomain/.style={rectangle, draw, fill=gray!20, rounded corners, align=center, minimum width=1cm, minimum height=1cm, font=\scriptsize},
    booldomain/.style={rectangle, draw, fill=green!30, align=center, minimum width=1cm, minimum height=0.5cm, font=\scriptsize},
    cartesian/.style={circle, draw, fill=blue!30, inner sep=0pt, minimum size=6mm},
    fields/.style={matrix of nodes, nodes={minimum height=3.8mm}, nodes in empty cells, ampersand replacement=\&},
    arrow/.style={single arrow, draw=black, minimum width=1.2cm},
]

\node(pic1) at (-5, -1) {
\begin{tikzpicture}
    
    \node (memorybank) [composedomain, minimum width=3.2cm, minimum height=3.3cm]  {};
    \node (pc) [signal, minimum width=1.3cm, minimum height=0.5cm, align=center, draw=black, font=\small, above left = 1.5cm and -0.5cm of memorybank] {\cref{irline:store}};

    \node (flags) [booldomain, minimum width=1.7cm, minimum height=0.5cm, above right = 0.9cm and -0.2cm of memorybank] at (memorybank.center) {};
    \node (base) [basedomain, minimum width=3.1cm, minimum height=1.1cm, above = 0.2cm of memorybank] {};
    \node (baseval) [align=center, font=\scriptsize, below = -0.2cm of base] at (base.center) {$sz\mapsto 2, i\mapsto 1,$ \\ $ p\mapsto \hex{2000}, \ldots$};
    \node (cache) [basedomain, minimum width=3cm, minimum height=0.7cm, above = -0.1cm of memorybank] at (memorybank.center) {};
    \node (storage) [basedomain, minimum width=3cm, minimum height=1.2cm, below = 0.2cm of cache] {};

    \node[below right] at (base.north west) {\small{$\scalarenv$}};
    \node(mbname)[below right] at (memorybank.north west) {\small{$\mb$}};
    \node[below right] at (cache.north west) {\tiny{$\cache$}};
    \node[below right] at (storage.north west) {\tiny{$\storage$}};
    \node[right] at (flags.west) {\tiny{$\objenv$}};

    \node(flagname) [above right = -.1cm and -1.15cm of flags, font=\tiny] { \ttfamily{used dirty}};
    \matrix(flagvalue)[fields, nodes={draw, anchor=center}, column sep=.09,row sep=.09, right = -.4cm of flags, font=\tiny] at (flags.center) { $\true$ \& $\true$ \\};
    \node(fieldname) [above right = -.1cm and -2.1cm of cache, font=\tiny] { \ttfamily{len \ind[1] cap \ind[1] buf}};
    \node(cacheaddr)[below right, font=\tiny, above left = -0.3cm and 0.6cm of cache] at (cache.center) {$\hex{1000}$};
    \matrix(cachefield)[fields, nodes={draw, anchor=center}, column sep=.09,row sep=.09, right = -.1cm of cache, font=\tiny] at (cacheaddr.east) { $\;0\;\;$ \& $\;1\;\;$ \& $\hex{100}$  \\};
    \node(addr1)[below right, font=\tiny, above left = 0.01cm and 0.6cm of storage] at (storage.center) {$\hex{1000}$};
    \matrix(storagefield1)[fields, nodes={draw, anchor=center}, column sep=.09,row sep=.09, right = -.1cm of storage, font=\tiny] at (addr1.east) { NA  \&  NA \& $\;$NA$\;\;$ \\};
    \matrix(storagefield2)[fields, nodes={draw, anchor=center}, column sep=.09,row sep=.09, below = -0.15cm of storagefield1, font=\tiny] { NA  \&  NA \& $\;$NA$\;\;$ \\};
    \node(addr2)[font=\tiny, left = .97cm of storagefield2] at (storagefield2.center) {$\hex{2000}$};
\end{tikzpicture}
};

\node(label1) at (-4.6, -3.9){(1)};

\node(step1)[arrow, align=center, dashed] at (-2, -1.5) {$\;$flush$\;$};

\node(pic2) at (1, -1) {
\begin{tikzpicture}
    
    \node (memorybank) [composedomain, minimum width=3.2cm, minimum height=3.3cm]  {};
    \node (pc) [signal, minimum width=1.3cm, minimum height=0.5cm, align=center, draw=black, font=\small, above left = 1.5cm and -0.5cm of memorybank] {\cref{irline:gep}};

    \node (flags) [booldomain, minimum width=1.7cm, minimum height=0.5cm, above right = 0.9cm and -0.2cm of memorybank] at (memorybank.center) {};
    \node (base) [basedomain, minimum width=3.1cm, minimum height=1.1cm, above = 0.2cm of memorybank] {};
    \node (baseval) [align=center, font=\scriptsize, below = -0.2cm of base] at (base.center) {$sz\mapsto 2, i\mapsto 1,$ \\ $ p\mapsto \hex{2000}, \ldots$};
    \node (cache) [basedomain, minimum width=3cm, minimum height=0.7cm, above = -0.1cm of memorybank] at (memorybank.center) {};
    \node (storage) [basedomain, minimum width=3cm, minimum height=1.2cm, below = 0.2cm of cache] {};

    \node[below right] at (base.north west) {\small{$\scalarenv$}};
    \node(mbname)[below right] at (memorybank.north west) {\small{$\mb$}};
    \node[below right] at (cache.north west) {\tiny{$\cache$}};
    \node[below right] at (storage.north west) {\tiny{$\storage$}};
    \node[right] at (flags.west) {\tiny{$\objenv$}};

    \node(flagname) [above right = -.1cm and -1.15cm of flags, font=\tiny] { \ttfamily{used dirty}};
    \matrix(flagvalue)[fields, nodes={draw, anchor=center}, column sep=.09,row sep=.09, right = -.4cm of flags, font=\tiny] at (flags.center) { $\false$ \& $\false$ \\};
    \node(fieldname) [above right = -.1cm and -2.1cm of cache, font=\tiny] { \ttfamily{len \ind[1] cap \ind[1] buf}};
    \node(cacheaddr)[below right, font=\tiny, above left = -0.3cm and 0.6cm of cache] at (cache.center) {$\hex{1000}$};
    \matrix(cachefield)[fields, nodes={draw, anchor=center}, column sep=.09,row sep=.09, right = -.1cm of cache, font=\tiny] at (cacheaddr.east) { $\;0\;\;$ \& $\;1\;$ \& $\hex{100}$  \\};
    \node(addr1)[below right, font=\tiny, above left = 0.01cm and 0.6cm of storage] at (storage.center) {$\hex{1000}$};
    \matrix(storagefield1)[fields, nodes={draw, anchor=center}, column sep=.09,row sep=.09, right = -.1cm of storage, font=\tiny] at (addr1.east) { $\;0\;\;$ \& $\;1\;$ \& $\hex{100}$  \\};
    \matrix(storagefield2)[fields, nodes={draw, anchor=center}, column sep=.09,row sep=.09, below = -0.15cm of storagefield1, font=\tiny] { NA  \&  NA \& $\;$NA$\;\;$ \\};
    \node(addr2)[font=\tiny, left = .97cm of storagefield2] at (storagefield2.center) {$\hex{2000}$};
    \draw[red, -{Stealth[fill=red]}] (cachefield.east) to[bend right,out=45,in=135] node [midway,right,pos=0.5,align=center, font=\small] {\ttfamily{flush}}  (storagefield1.east);
\end{tikzpicture}
};

\node(label2) at (0.9, -3.9){(2)};
\node(step2)[arrow, align=center, shape border rotate=270, dashed] at (0.9, -4.5) {$\;$\\update};

\node(pic3) at (1, -8) {
\begin{tikzpicture}
    
    \node (memorybank) [composedomain, minimum width=3.2cm, minimum height=3.3cm]  {};
    \node (pc) [signal, minimum width=1.3cm, minimum height=0.5cm, align=center, draw=black, font=\small, above left = 1.5cm and -0.5cm of memorybank] {\cref{irline:gep}};

    \node (flags) [booldomain, minimum width=1.7cm, minimum height=0.5cm, above right = 0.9cm and -0.2cm of memorybank] at (memorybank.center) {};
    \node (base) [basedomain, minimum width=3.1cm, minimum height=1.1cm, above = 0.2cm of memorybank] {};
    \node (baseval) [align=center, font=\scriptsize, below = -0.2cm of base] at (base.center) {$sz\mapsto 2, i\mapsto 1,$ \\ $ p\mapsto \hex{2000}, \ldots$};
    \node (cache) [basedomain, minimum width=3cm, minimum height=0.7cm, above = -0.1cm of memorybank] at (memorybank.center) {};
    \node (storage) [basedomain, minimum width=3cm, minimum height=1.2cm, below = 0.2cm of cache] {};

    \node[below right] at (base.north west) {\small{$\scalarenv$}};
    \node(mbname)[below right] at (memorybank.north west) {\small{$\mb$}};
    \node[below right] at (cache.north west) {\tiny{$\cache$}};
    \node[below right] at (storage.north west) {\tiny{$\storage$}};
    \node[right] at (flags.west) {\tiny{$\objenv$}};

    \node(flagname) [above right = -.1cm and -1.15cm of flags, font=\tiny] { \ttfamily{used dirty}};
    \matrix(flagvalue)[fields, nodes={draw, anchor=center}, column sep=.09,row sep=.09, right = -.4cm of flags, font=\tiny] at (flags.center) { $\true$ \& $\false$ \\};
    \node(fieldname) [above right = -.1cm and -2.1cm of cache, font=\tiny] { \ttfamily{len \ind[1] cap \ind[1] buf}};
    \node(cacheaddr)[below right, font=\tiny, above left = -0.3cm and 0.6cm of cache] at (cache.center) {$\hex{2000}$};
    \matrix(cachefield)[fields, nodes={draw, anchor=center}, column sep=.09,row sep=.09, right = -.1cm of cache, font=\tiny] at (cacheaddr.east) { NA  \&  NA \& $\;$NA$\;$$\;$ \\};
    \node(addr1)[below right, font=\tiny, above left = 0.01cm and 0.6cm of storage] at (storage.center) {$\hex{1000}$};
    \matrix(storagefield1)[fields, nodes={draw, anchor=center}, column sep=.09,row sep=.09, right = -.1cm of storage, font=\tiny] at (addr1.east) { $\;0\;\;$ \& $\;1\;$ \& $\hex{100}$  \\};
    \matrix(storagefield2)[fields, nodes={draw, anchor=center}, column sep=.09,row sep=.09, below = -0.15cm of storagefield1, font=\tiny] { NA  \&  NA \& $\;$NA$\;\;$ \\};
    \node(addr2)[font=\tiny, left = .97cm of storagefield2] at (storagefield2.center) {$\hex{2000}$};
    \draw[red, {Stealth[fill=red]}-] (cachefield.east) to[bend right,out=45,in=135] node [midway,right,pos=0.5,align=center, font=\small] {\ttfamily{update}}  (storagefield2.east);
\end{tikzpicture}
};

\node(label3) at (0.9, -10.9){(3)};
\node(step3)[arrow, align=center, shape border rotate=180, dashed] at (-2, -8.5) {$\;$store$\;$};

\node(pic4) at (-5, -8) {
\begin{tikzpicture}
    
    \node (memorybank) [composedomain, minimum width=3.2cm, minimum height=3.3cm]  {};
    \node (pc) [signal, minimum width=1.3cm, minimum height=0.5cm, align=center, draw=black, font=\small, above left = 1.5cm and -0.5cm of memorybank] {\cref{irline:gep}};

    \node (flags) [booldomain, minimum width=1.7cm, minimum height=0.5cm, above right = 0.9cm and -0.2cm of memorybank] at (memorybank.center) {};
    \node (base) [basedomain, minimum width=3.1cm, minimum height=1.1cm, above = 0.2cm of memorybank] {};
    \node (baseval) [align=center, font=\scriptsize, below = -0.2cm of base] at (base.center) {$sz\mapsto 2, i\mapsto 1,$ \\ $ p\mapsto \hex{2000}, \ldots$};
    \node (cache) [basedomain, minimum width=3cm, minimum height=0.7cm, above = -0.1cm of memorybank] at (memorybank.center) {};
    \node (storage) [basedomain, minimum width=3cm, minimum height=1.2cm, below = 0.2cm of cache] {};

    \node[below right] at (base.north west) {\small{$\scalarenv$}};
    \node(mbname)[below right] at (memorybank.north west) {\small{$\mb$}};
    \node[below right] at (cache.north west) {\tiny{$\cache$}};
    \node[below right] at (storage.north west) {\tiny{$\storage$}};
    \node[right] at (flags.west) {\tiny{$\objenv$}};

    \node(flagname) [above right = -.1cm and -1.15cm of flags, font=\tiny] { \ttfamily{used dirty}};
    \matrix(flagvalue)[fields, nodes={draw, anchor=center}, column sep=.09,row sep=.09, right = -.4cm of flags, font=\tiny] at (flags.center) { $\true$ \& $\true$ \\};
    \node(fieldname) [above right = -.1cm and -2.1cm of cache, font=\tiny] { \ttfamily{len \ind[1] cap \ind[1] buf}};
    \node(cacheaddr)[below right, font=\tiny, above left = -0.3cm and 0.6cm of cache] at (cache.center) {$\hex{2000}$};
    \matrix(cachefield)[fields, nodes={draw, anchor=center}, column sep=.09,row sep=.09, right = -.1cm of cache, font=\tiny] at (cacheaddr.east) { $\color{red}\;1\;$ \& $\;$NA$\;$ \& $\;$NA$\;$  \\};
    \node(addr1)[below right, font=\tiny, above left = 0.01cm and 0.6cm of storage] at (storage.center) {$\hex{1000}$};
    \matrix(storagefield1)[fields, nodes={draw, anchor=center}, column sep=.09,row sep=.09, right = -.1cm of storage, font=\tiny] at (addr1.east) { $\;0\;\;$ \& $\;1\;$ \& $\hex{100}$  \\};
    \matrix(storagefield2)[fields, nodes={draw, anchor=center}, column sep=.09,row sep=.09, below = -0.15cm of storagefield1, font=\tiny] { NA  \&  NA \& $\;$NA$\;\;$ \\};
    \node(addr2)[font=\tiny, left = .97cm of storagefield2] at (storagefield2.center) {$\hex{2000}$};
\end{tikzpicture}
};

\node(label4) at (-4.6, -10.9){(4)};

\node(step)[arrow, align=center, shape border rotate=270] at (-4.6, -4.5) {$\;$\\$\;\;$step$\;\;$};

\end{tikzpicture}}
\subcaption{} 
\label{subfig:rumm-details}
\end{subfigure}
\end{minipage}
    \caption{(a) A program, and (b) an  execution of \cref{irline:store} under \memModel. } 
    \label{fig:rumm-example}
\end{figure}

A memory model defines how memory is structured and accessed in the operational
semantics (i.e., execution) of the program. The standard C memory model (CMM)
treats each allocation as a blob of bytes. Specifically, each memory object is a
blob of bytes (logically sub-divided into fields). A pointer is a pair $(b, o)$
of an object identifier $b$ (a.k.a., the \emph{base} address) and a numeric
offset $o$ within that object. At allocation, an object occupies a blob in
memory at an address determined by the memory allocator. Each memory operation
is performed through a pointer to access the object's content. In practice, CMM
is typically implemented by a flat memory model of the underlying architecture.
However, non-flat memory models with multiple address spaces are common,
especially in embedded systems~\cite{DBLP:conf/afips/Huston82a}.

In this paper, we introduce a new memory model, called \emph{recent-use memory
  model} (\memModel ), that differentiates between the most recently used (MRU)
object and other memory objects. \memModel partitions memory into multiple
\emph{banks}, each with (a) a \emph{storage} -- a blob of bytes that
permanently stores memory objects, and (b) a \emph{cache} -- a blob of bytes
that temporarily holds the MRU object of that bank. The notion of objects and
pointers in \memModel is exactly as in CMM. Furthermore, \memModel is
parameterized by a function $\mlocate$ that maps allocation sites to specific
memory banks of \memModel. This is similar to a pool allocation, where objects are
allocated in different pools~\cite{DBLP:conf/pldi/LattnerA05}. Each object is
allocated as a blob in the selected bank’s storage, with each bank managing its allocations.

What makes \memModel special is its handling of read and write operations. To access
an object $x$ from a given bank, $x$ is first loaded into the cache and then
accessed from there. If a different object $y$ currently occupies the cache, $y$ is
flushed back to its place in its memory bank before $x$ is loaded. Thus,
multiple read and write operations that work on the same object only use the
cache, until the cache is flushed when a new object, from the same bank, is
accessed.

\cref{subfig:cfg} shows a \crabir for the \code{for} loop in
\cref{fig:discuss-example}. Variables prefixed with \code{@} are the fields of
\code{byte\_buf}. The loop starts at the \emph{entry block} and checks whether
the counter \code{i} meets the enter/exit condition. In \crabir, \code{assume}
is used to enforce this condition. The loop initializes a memory object,
increments the counter, and loops back to the loop entry. \cref{subfig:rumm-details}
illustrates the execution of \cref{irline:store} during the \emph{second}
iteration of the loop.
\cref{subfig:rumm-details}(1) shows the state at \cref{irline:store}, where
scalar variables map to their values as $\scalarenv$ and a memory bank $\mb$ is
provided to store memory objects allocated at \cref{irline:alloc}. We assume the
first two iterations allocate objects at addresses \hex{1000} and \hex{2000},
respectively. The fields of each object are visually represented as slots, with
either concrete values or marked as not available (NA) if uninitialized. The
storage keeps two uninitialized objects, while the cache holds the MRU object.
The object at address \hex{1000} is the MRU since its last access is at
\cref{irline:oldstore} during the first iteration. The cache status is indicated
by two flags: \emph{used}, indicating the cache is active, and \emph{dirty},
meaning the cache value has been updated. When \code{store} at
\cref{irline:store} accesses the object with address \hex{2000}, the cache
flushes the object (\hex{1000}) back to the storage
(\cref{subfig:rumm-details}(2)) and updates with the uninitialized object from
the storage (\cref{subfig:rumm-details}(3)). The cache is then ready to write
\code{@len} with a value of $1$ (\cref{subfig:rumm-details}(4)).

A cache in each bank is crucial because it temporarily isolates the accessed object from others. \cref{sec:abs-sem} describes our abstraction based on allocation-site abstraction, which collapses all objects in a bank into a single summary object. Without a cache, updates must be weak to ensure correctness, inevitably reducing precision of analysis. With a cache, however, write operations can be modeled as strong updates after abstraction. For this reason, \memModel incorporates caches to improve precision.

We argue that \memModel is compatible with CMM --  there is a bisimulation between \memModel and CMM semantics. This follows from: (1) \memModel organizes memory objects into separate, non-overlapping memory banks; (2) The usage of cache is an extra step that does not invalidate the properties of each object. The semantics of \crabir are the same under both memory models. In the following, we formalize the concrete semantics of \crabir under \memModel.

\setlength{\columnsep}{-1cm}
\setlength{\columnseprule}{-5mm}
\begin{figure}[t]
\hspace{-1em}
\begin{minipage}[t]{0.47\textwidth}
\begin{algorithmic}
\MyFunction{$\tran{\varptr := \makeref(\varfld, \varint)}{\text{\memModel}}$}{$\stat$}
\State $\cassign{\langle\scalarenv, \memory\rangle}{\stat}$
\State $\cassign{\mb}{\mlocate(\varfld, \memory)}$
\State $\cassign{\langle \cache, \storage, \objenv\rangle}{\mb}$
\State $\cassign{\langle\_,\sz\rangle}{\scalarenv[\varint]}$
\State $\Tlet \langle\varptrbase, \storage'\rangle \gets$
\Statex $\ind[4]\allocator_{\mb}(\storage, \sz)\Tin$
\State $\Tlet \scalarenv' \gets$
\Statex $\ind[4]\scalarenv[\varptr \mapsto \langle \varptrbase, 0 \rangle]\Tin$
\State $\cassign{\mb'}{\langle \cache, \storage', \objenv\rangle}$
\State $\langle\scalarenv', \memory \setminus \{\mb\} \cup \{\mb'\}\rangle$
\EndMyFunction
\State
\MyFunction{$\tran{\varscalar := \loadref(\varptr, \varfld)}{\text{\memModel}}$}{$\stat$}
\State $\cassign{\langle\scalarenv, \memory\rangle}{\stat}$
\State $\cassign{\mb}{\mlocate(\varfld, \memory)}$
\State $\cassign{\langle \varptrbase, \_ \rangle}{\scalarenv[\varptr]}$
\State $\cassign{\mb'}{\invalidateCacheIfMiss(\mb, \varptrbase)}$
\State $\cassign{\langle \cache, \_, \_ \rangle}{\mb'}$
\State $\cassign{\langle\_,\fldenv\rangle}{\cache}$
\State $\Tlet \scalarenv' \gets$
\Statex $\ind[4]\scalarenv[\varscalar \mapsto \fldenv[\varfld]]\Tin$
\State $\langle\scalarenv', \memory \setminus \{\mb\} \cup \{\mb'\}\rangle$
\EndMyFunction
\end{algorithmic}
\end{minipage}
\hspace{-.4em}
\vline
\hspace{-.8em}
\begin{minipage}[t]{0.56\textwidth}
\begin{algorithmic}
\MyFunction{$\tran{\varptrtwo, \varfldtwo := \gepref(\varptrone, \varfldone, \varint)}{\text{\memModel}}$}{$\stat$}
\State $\cassign{\langle\scalarenv, \memory\rangle}{\stat}$
\State $\cassign{\langle \varptronebase, \offset\rangle}{\scalarenv[\varptr]}$
\State $\cassign{\langle\_,\Tval\rangle}{\scalarenv[\varint]}$
\State $\cassign{\offset'}{\offset + \Tval}$
\State $\Tlet \scalarenv' \gets$
\Statex $\ind[2]\scalarenv[\varptrtwo \mapsto \langle \varptronebase, \offset' \rangle]\Tin$
\State $\langle\scalarenv', \memory\rangle$
\EndMyFunction
\State
\MyFunction{$\tran{\storeref(\varptr, \varfld, \varscalar)}{\text{\memModel}}$}{$\stat$}
\State $\cassign{\langle\scalarenv, \memory\rangle}{\stat}$
\State $\cassign{\mb}{\mlocate(\varfld, \memory)}$
\State $\cassign{\langle \varptrbase, \_ \rangle}{\scalarenv[\varptr]}$
\State $\cassign{\mb'}{\invalidateCacheIfMiss(\mb, \varptrbase)}$
\State $\cassign{\langle\cache, \storage, \_ \rangle}{\mb'}$
\State $\cassign{\langle\cachebase,\fldenv\rangle}{\cache}$
\State $\Tlet \cache' \gets \langle\cachebase,$
\Statex $\ind[4] \fldenv[\varfld \mapsto \scalarenv[\varscalar]]\rangle\Tin$
\State $\Tlet \mb'' \gets$
\Statex $\ind[4]\langle  \cache', \storage, \langle\true, \true\rangle \rangle\Tin$
\State $\langle\scalarenv, \memory \setminus \{ \mb \} \cup \{\mb^{''}\}\rangle$
\EndMyFunction
\end{algorithmic}
\end{minipage}
\caption{\crabir statements operating under \memModel.}
\label{algo:transformer_concrete}
\vspace{-1em}
\end{figure}
\begin{figure}[t]
\begin{minipage}[t]{0.71\textwidth}
\begin{algorithmic}
\MyFunction{$\invalidateCacheIfMiss$}{$\mb, \varptrbase$}
\State $\cassign{\langle \cache, \storage, \langle\cacheused, \cachedirty\rangle\rangle}{\mb}$
\State $\cassign{\langle\cachebase,~\_\rangle}{\cache}$
\Let{$\mb'$}
    {\If{$\neg \cacheused \land \varptrbase \neq \cachebase$}
    \State $\Tlet\storage' \gets \mathbf{if}~ \cachedirty ~\mathbf{then}~ $
    \Statex $\ind[5] \lang{flush}(\cache, \storage) ~\mathbf{else}~ \storage\Tin$
    \State $\cassign{\cache'}{\lang{refresh}({\storage}^{\prime}, \varptrbase)}$
    \State $\Tlet \mb' \gets $
    \Statex $\ind[5] \langle \cache', \storage', \langle\true, \false\rangle\rangle\Tin$
    \State $\mb'$
    \Else $~\mb$
    \EndIf}
\EndLet $\mb'$ 
\EndMyFunction
\end{algorithmic}
\end{minipage}
\hspace{-4em}
\vline
\begin{minipage}[t]{0.38\textwidth}
\begin{algorithmic}
\MyFunction{$\lang{flush}$}{$\cache, \storage$}
\State $\Tlet\langle\cachebase,\fldenv\rangle \gets $
\Statex $\ind[4] \cache\Tin$
\State $\storage[\cachebase \mapsto \fldenv]$
\EndMyFunction
\State
\MyFunction{$\lang{refresh}$}{$\storage, \varptrbase$}
\State $\langle\varptrbase, \storage[\varptrbase]\rangle$
\EndMyFunction
\end{algorithmic}
\end{minipage}
\caption{Cache operations.}
\label{algo:cache_concrete}
\vspace{-1em}
\end{figure}

A \crabir program has scalars (i.e., integers $\vint$ and pointers $\vptr$)
whose values are represented as \emph{cells}. A cell, $\cell \in
\Cell:\mathbb{N} \times \mathbb{Z}$, represents either a pointer's base address
and offset, denoted as $\langle \baseaddr, \offset \rangle$, or an integer
value: $\langle 0, \Tval\rangle$. Formally, a scalar state is $\scalarenv\in
\ScalarEnv : \vscalar \mapsto \Cell$. To avoid redundancy, we explicitly associate the
base address of a $\varptr$ with a ghost variable $\varptrbase\in \vptrbase$.
For example, if a pointer \code{p} is $\langle\hex{100}, 8\rangle$, then
$\mathtt{p}^\mathtt{base}$ is $\hex{100}$.

The memory is modeled as a set of memory banks, $\memory \in \Memory :\{\mb ~|~
\mb \allowbreak \in \MemoryBank\}$. Each bank, $\mb\in \MemoryBank : \Cache \times \Storage
\times \ObjEnv$, holds memory values for cache, storage, and boolean flags. The
cache, $\cache\in\Cache : \mathbb{N} \times \FldEnv$, includes the cached
object's base address (as $\cachebase$) and field values. 
The field values (as
cells) are kept in an environment $\fldenv\in \FldEnv: \vfld \mapsto \Cell$. The
storage, $\storage\in \Storage: \mathbb{N} \mapsto \FldEnv$, maps base addresses
of memory objects to the corresponding field environment. The cache boolean
flags, $\objenv$, indicate if it is occupied ($\cacheused$) and overwritten
($\cachedirty$). Overall, a concrete program state $\stat \in \cstat$ is a
tuple: $\langle \scalarenv, \mem \rangle$. We assume $\mlocate$ maps a field
variable and memory state to a memory bank, indicating in which bank the field
is stored.

\cref{algo:transformer_concrete,algo:cache_concrete} describe the changes to a program state at each memory and pointer arithmetic statements in \crabir. The function $\tran{\cdot}{\text{\memModel}}(\cdot)$ takes a statement and a program state and returns the computed state under \memModel. The initial state's $\scalarenv$ is an empty map. Each bank $\mb$ contains an empty $\cache$, an empty map $\storage$, and a $\langle \false, \false \rangle$ cache flags.

The $\makeref$ statement creates a new memory object of size $\varint$, assigns it to a specific bank's storage. The bank is determined by $\varfld$ through $\mlocate$, and its $\allocator$ constructs the object and returns its base address assigned to $\varptr$.

The $\gepref$ computes a new pointer value for $\varptrtwo$ by adding an offset $\varint$ to the pointer value of $\varptrone$. Earlier, we assume all pointer arithmetic stays inbounds, so the $\varptrtwo$ and $\varptrone$ have the same base address but (presumably) different offsets. 

The $\loadref$ operation accesses the object pointed by $\varptr$ from the cache associated with the corresponding memory bank. To ensure the object is cached, we use the $\invalidateCacheIfMiss$ function to check if the cache is missed. If so, we flush the cache back to the storage with \code{flush} if the cache is modified, and then load the new MRU object by calling \code{refresh}. The \code{flush} function moves the currently cached object into $\storage$, while \code{refresh} refreshes the cache with the object pointed by $\varptr$. After that, the object at $\varptr$ is in the cache, so the flag $\cacheused$ is set to \code{true}. The value of $\varscalar$ in $\scalarenv$ gets updated by the cached field $\varfld$. Similarly, $\storeref$ updates the field for the object, using $\invalidateCacheIfMiss$ to ensure it is in the cache. The flag $\cachedirty$ is set to \code{true}, indicating the object has been modified.

Overall, the \memModel offers a different way to organize C memory by partitioning it into multiple banks, with additional space to temporarily hold a memory object for reads and writes. This setup is very convenient because it provides a straightforward memory abstraction by summarizing all objects from the same bank into one and simplifies the design of \mrudomain, as described in \cref{sec:abs-sem}.

\section{An Abstract Domain for Inferring Object Invariants}
\label{sec:abs-sem}

In this section, we introduce \mrudomain, a new abstract domain that is a (partially) reduced product
of the domains for scalars, pointers, and objects. After setting up the domain,
we detail key transfer functions and the reduction procedure.

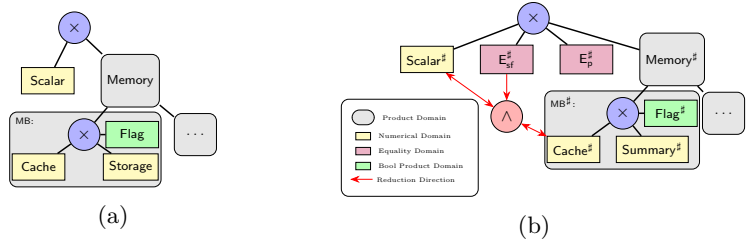
\begin{figure}[t]
\begin{minipage}[c]{0.45\textwidth}\centering%
\begin{subfigure}{\textwidth}\centering%
\scalebox{0.7}{
\begin{tikzpicture}[
    basedomain/.style={rectangle, draw, fill=yellow!30, align=center, minimum width=1cm, minimum height=0.5cm, font=\scriptsize},
    eqdomain/.style={rectangle, draw, fill=purple!30, align=center, minimum width=1cm, minimum height=0.5cm, font=\scriptsize},
    booldomain/.style={rectangle, draw, fill=green!30, align=center, minimum width=1cm, minimum height=0.5cm, font=\scriptsize},
    composedomain/.style={rectangle, draw, fill=gray!20, rounded corners, align=center, minimum width=1cm, minimum height=1cm, font=\scriptsize},
    cartesian/.style={circle, draw, fill=blue!30, inner sep=0pt, minimum size=6mm},
    reduce/.style={circle, draw, fill=red!30, inner sep=0pt, minimum size=6mm},
    background/.style={rectangle, draw, fill=white, rounded corners, inner sep=0.2cm},
    myarrow/.style={thick},
    reducearrow/.style={thick, fill=red}
]

\node[basedomain] (basedom) {$\ScalarEnv$};
\node[composedomain, right=0.5 cm of basedom] (odimap) {$\Memory$};
\node[cartesian, above left=0.4 cm of odimap] (cross) {\(\times\)};
\node[cartesian, below left=0.3cm and 0.1cm of odimap] (cross3) {\(\times\)};
\node[composedomain, minimum width=0.8cm, minimum height=0.8cm, right=1.4 cm of cross3] (dots) {\(\ldots\)};
\node[basedomain, below left=0.2 cm of cross3] (cache) {$\Cache$};
\node[basedomain, below right=0.2 cm of cross3] (sum) {$\Storage$};
\node[booldomain, right=0.1 cm of cross3] (flag) {$\ObjEnv$};

\draw[myarrow] (cross) -- (basedom);
\draw[myarrow] (cross) -- (odimap);
\draw[myarrow] (cross3) -- (odimap);
\draw[myarrow] (dots) -- (odimap);
\draw[myarrow] (cross3) -- (cache);
\draw[myarrow] (cross3) -- (sum);
\draw[myarrow] (cross3) -- (flag);

\begin{scope}[on background layer]
\node[background, fill=gray!20, fit=(cross3) (cache) (sum) (flag), scale = 0.9] (memorybank) {};
\end{scope}

\node[below right, font=\tiny] at (memorybank.north west) {$\MemoryBank$:};

\end{tikzpicture}
}
  \caption{}
  \label{subfig:credom}
\end{subfigure}
\end{minipage}
\begin{minipage}[c]{0.45\textwidth}\centering%
\begin{subfigure}{\textwidth}\centering%
\scalebox{0.7}{
\begin{tikzpicture}[
    basedomain/.style={rectangle, draw, fill=yellow!30, align=center, minimum width=1cm, minimum height=0.5cm, font=\scriptsize},
    eqdomain/.style={rectangle, draw, fill=purple!30, align=center, minimum width=1cm, minimum height=0.5cm, font=\scriptsize},
    booldomain/.style={rectangle, draw, fill=green!30, align=center, minimum width=1cm, minimum height=0.5cm, font=\scriptsize},
    composedomain/.style={rectangle, draw, fill=gray!20, rounded corners, align=center, minimum width=1cm, minimum height=1cm, font=\scriptsize},
    cartesian/.style={circle, draw, fill=blue!30, inner sep=0pt, minimum size=6mm},
    reduce/.style={circle, draw, fill=red!30, inner sep=0pt, minimum size=6mm},
    background/.style={rectangle, draw, fill=white, rounded corners, inner sep=0.2cm},
    myarrow/.style={thick},
    reducearrow/.style={thick, fill=red}
]

\node[basedomain] (basedom) {$\Basedom$};
\node[eqdomain, right=0.5cm of basedom] (eqfieldscalar) {$\eqsfdom$};
\node[eqdomain, right=0.5cm of eqfieldscalar] (addrsdom) {$\AddrsDom$};
\node[composedomain, right=0.5 cm of addrsdom] (odimap) {$\AbsMemory$};
\node[cartesian, above left=0.4 cm of addrsdom] (cross) {\(\times\)};
\node[cartesian, below left=0.3cm and 0.1cm of odimap] (cross3) {\(\times\)};
\node[composedomain, minimum width=0.8cm, minimum height=0.8cm, right=1.2 cm of cross3] (dots) {\(\ldots\)};
\node[basedomain, below left=0.3 cm of cross3] (cache) {$\AbsCache$};
\node[basedomain, right=0.3 cm of cache] (sum) {$\AbsSummary$};
\node[booldomain, right=0.1 cm of cross3] (flag) {$\AbsObjEnv$};

\node[reduce, below=0.5cm of eqfieldscalar] (reduce) {\(\land\)};

\draw[myarrow] (cross) -- (basedom);
\draw[myarrow] (cross) -- (addrsdom);
\draw[myarrow] (cross) -- (odimap);
\draw[myarrow] (cross) -- (eqfieldscalar);
\draw[myarrow] (cross3) -- (odimap);
\draw[myarrow] (dots) -- (odimap);
\draw[myarrow] (cross3) -- (cache);
\draw[myarrow] (cross3) -- (sum);
\draw[myarrow] (cross3) -- (flag);
\draw[red, {Stealth[fill=red]}-] (reduce) -- (eqfieldscalar);
\draw[red,{Stealth[fill=red]}-{Stealth[fill=red]}] (reduce) -- (cache);
\draw[red,{Stealth[fill=red]}-{Stealth[fill=red]}] (reduce) -- (basedom);

\begin{scope}[scale=0.7, every node/.append style={transform shape}]
\node[rectangle, draw, fill=gray!20, rounded corners, minimum width=0.6cm, minimum height=0.4cm, below left =1 cm and 0.7cm of basedom] (bg) {};
\node[right=0.1cm of bg, font=\tiny] (pd) {Product Domain};
\node[basedomain, below =0.2 cm of bg, scale=0.4] (yellow) {};
\node[right=0.1cm of yellow, font=\tiny] (dyellow) {Numerical Domain};
\node[eqdomain, below =0.2 cm of yellow, scale=0.4] (purple) {};
\node[right=0.1cm of purple, font=\tiny] (dpurple) {Equality Domain};
\node[booldomain, below =0.2 cm of purple, scale=0.4] (green) {};
\node[right=0.1cm of green, font=\tiny] (dgreen) {Bool Product Domain};
\node[below left =0.2 cm of green, font=\tiny] (empty) {$\quad$};
\node[right=0.6 cm of empty, font=\tiny] (empty2) {Reduction Direction};
\draw[red, {Stealth[fill=red]}-] (empty) -- (empty2);
\end{scope}

\begin{scope}[on background layer]
\node[background, fill=gray!20, fit=(cross3) (cache) (sum) (flag), scale = 0.9] (memorybank) {};
\end{scope}

\node[below right, font=\tiny] at (memorybank.north west) {$\AbsMemoryBank$:};

\begin{scope}[on background layer]
\node[background, fit=(bg) (pd) (yellow) (dyellow) (green) (dgreen) (purple) (purple) (empty2)] {};
\end{scope}
\end{tikzpicture}
}
  \caption{}
  \label{subfig:absdom}
\end{subfigure}
\end{minipage}
\caption{(a) Concrete domain and (b) \mrudomain hierarchy.}
\label{fig:domain_hierarchy}
\vspace{-1em}
\end{figure}

\begin{figure}[t]
\begin{minipage}[b]{0.48\textwidth}\centering%
{
\tiny
\begin{alignat*}{2}
\scalarenv &\in \Basedom \; &&{\Def= } \;  \numdom(\vscalar) \\
\regfldseq &\in \eqsfdom \; &&{\Def= } \;  \eqdom(\vscalar\cup \vfld) \\
\addrseq  &\in \AddrsDom  \; &&{\Def= } \; \eqdom(\vptrbase) \\
\langle\cacheused, \cachedirty, \cacheinitpack \rangle &\in \AbsObjEnv \; &&{\Def= } \; \pair{\pair{\booldom}{\booldom}}{\booldom} \\
\abscache &\in \AbsCache \; &&{\Def= } \;  \numdom(\vfld)\\
\abssummary &\in \AbsSummary \; &&{\Def= } \; \numdom(\vfld) \\
\mb &\in \AbsMemoryBank  \; &&{\Def= } \; \AbsCache \times \AbsSummary \times \AbsObjEnv \\
\absmemory &\in \AbsMemory \; &&{\Def= } \; \prod_i \{ \AbsMemoryBank_i \} \\
\absstat &\in \astat \; &&{\Def= } \; \Basedom \times \AbsMemoryBank \times \AddrsDom \times \eqsfdom
\end{alignat*}
}
\vspace{-2em}
\caption{Abstract semantic domains.}
\label{fig:abs_domains}
\end{minipage}
\begin{minipage}[b]{0.48\textwidth}\centering%
\scalebox{0.63}{
  \begin{tikzpicture}[
    basedomain/.style={rectangle, draw, fill=yellow!30, align=center, minimum width=1cm, minimum height=0.5cm, font=\scriptsize},
    eqdomain/.style={rectangle, draw, fill=purple!30, align=center, minimum width=1cm, minimum height=0.5cm, font=\scriptsize},
    booldomain/.style={rectangle, draw, fill=green!30, align=center, minimum width=1cm, minimum height=0.5cm, font=\scriptsize},
    composedomain/.style={rectangle, draw, fill=gray!20, rounded corners, align=center, minimum width=1cm, minimum height=1cm, font=\scriptsize},
    cartesian/.style={circle, draw, fill=blue!30, inner sep=0pt, minimum size=6mm},
    reduce/.style={circle, draw, fill=red!30, inner sep=0pt, minimum size=6mm},
    fields/.style={matrix of nodes, nodes in empty cells, ampersand replacement=\&},
    background/.style={rectangle, draw, fill=gray!10, rounded corners, inner sep=0.2cm},
    myarrow/.style={thick},
    reducearrow/.style={thick, fill=red}
]
    
    \node (odimap) [composedomain, minimum width=2.8cm, minimum height=3.6cm]  {};
    \node (pc) [signal, minimum width=1.3cm, minimum height=0.5cm, align=center, draw=black, font=\small, above left = 1.5cm and -0.5cm of odimap] {\cref{irline:absstate}};
    \node (basedom) [basedomain, minimum width=2.8cm, minimum height=1cm, above = 0.2cm of odimap] {$\mathit{sz} =2 \land \mathit{i} = 1 \land \ldots$};
    \node (flags) [booldomain, minimum width=2.05cm, minimum height=0.5cm, above right = 1cm and -.8cm of odimap] at (odimap.center) {};
    \node(flagname) [above right = -.1cm and -1.7cm of flags, font=\tiny] { \ttfamily{used dirty ispack} };
    \matrix(flagvalue)[fields, nodes={draw, anchor=center}, column sep=-0.7,row sep=.5, right = -.7cm of flags, font=\tiny] at (flags.center) { $\true$ \& $\true$ \& $\true$ \\};

    \node (cache) [basedomain, minimum width=2.5cm, minimum height=1cm, above = -0.1cm of odimap] at (odimap.center) {$\ldots$};
    \node (cache2) [basedomain, minimum width=2.5cm, minimum height=1.5cm, right = 1.3cm of cache, text=red] {$\mathit{cap} = 2 $\\$ \mathit{len} = 1 ~\land$\\$\ldots$};
    \node (summary) [basedomain, minimum width=2.5cm, minimum height=1.5cm, below = 0.1cm of cache] {$\mathit{cap} = 1  ~\land  $\\$ \mathit{len} = 0 ~\land$\\$\ldots$};
    \node (eqfldscldom) [eqdomain, minimum width=2.5cm, minimum height=0.8cm, below right = -0.85cm and 0.2cm of basedom] {$\mathit{i} \approx \mathit{len} \land \mathit{sz} \approx \mathit{cap}$};
    \node (eqptrdom) [eqdomain, minimum width=2.5cm, minimum height=0.8cm, above = 0.2cm of eqfldscldom] {$p^\mathit{base} \approx \cachebase\land\ldots$};

    \node[below right] at (basedom.north west) {\small{$\scalarenv$}};
    \node[below right] at (odimap.north west) {\small{$\mb$}};
    \node[below right] at (eqfldscldom.north west) {\tiny{$\regfldseq$}};
    \node[below right] at (eqptrdom.north west) {\tiny{$\addrseq$}};
    \node[below right] at (cache.north west) {\tiny{$\cache$}};
    \node[below right] at (cache2.north west) {\tiny{$\cache$}};
    \node[below right] at (summary.north west) {\tiny{$\summary$}};
    \node[right = -.05cm of flags] at (flags.west) {\tiny{$\absobjenv$}};

    \draw[red, {-{Stealth[fill=red]}}] (cache) -- node [midway,align=center] {after \\reduce} (cache2);
 \end{tikzpicture}}
\caption{State at \cref{irline:absstate}, 2nd iteration.}
\label{fig:absstate-emp}
\end{minipage}
\vspace{-1em}
\end{figure}

Similar to the concrete domain in \cref{subfig:credom}, the \mrudomain is shown
in~\cref{subfig:absdom}. It is a reduced product of four domains: (a) a
numerical domain $\Basedom$, (b) an equality domain $\AddrsDom$, (c) an equality
domain $\eqsfdom$, and (d) a collection of product domains $\AbsMemory :
\{\AbsMemoryBank\}$. $\AbsMemoryBank$ is a product of two numerical domains, and
three Boolean domains: $\AbsCache \times \AbsSummary \times \AbsObjEnv$. 
\cref{fig:abs_domains} shows the abstract semantic domains where variables are mapped to unique \emph{dimensions} of each abstract domain. 
Most domains correspond to those in concrete semantics, except for a few that provide additional information.
Specifically, $\eqsfdom$ represents the value equivalence of fields and scalars, which enables information propagation between $\Basedom$ and $\AbsCache$ for domain reduction. $\AddrsDom$ captures the aliasing properties of pointers, indicating which pointer refers to which object.
The added Boolean domain in $\AbsObjEnv$ is a flag for later use.
All domains are parameterized by relational abstract domains like Zones~\cite{DBLP:conf/pado/Mine01}. An abstract state $\absstat$ is represented by lattice elements within the \mrudomain.

\cref{fig:absstate-emp} shows the abstract state at \cref{irline:absstate}
during the second iteration of the \crabir example from \cref{subfig:cfg}. We
assume that the Zones domain is used for equality and numerical domains. We only
show the invariants for scalars \code{i} and \code{sz}, and fields \code{len}
and \code{cap}. $\scalarenv$ shows invariants for the scalars \code{i} and
\code{sz}. The sole memory bank $mb$ represents the objects of type
\code{byte\_buf}. The $\abscache$ shows the invariants for the MRU
\code{byte\_buf} object referenced by pointer~\code{p}. This
follows from the equality $p^\mathit{base} \approx \cachebase$ in
$\addrseq$. The $\abscache$ does not have any explicit invariants for fields.
However, the fields invariants are \emph{implicitly} represented through the
invariants in $\scalarenv$ and the equalities in $\regfldseq$, $i \approx
\mathit{len}$ and $\mathit{sz} \approx \mathit{cap}$, that connect fields and
scalars. These equalities are established during field writes. For instance,
$\mathit{i} \approx \mathit{len}$ is there because instruction \code{store(@len,
  plen, i)} was used to update the field \code{len} with scalar \code{i}.
Finally, $\abssummary$ shows the object invariants for the objects initialized
at the first iteration. Specifically, the fields of that object satisfy
\code{len <= cap}.

\begin{figure}[t]
\hspace{-1em}
\begin{minipage}[t]{0.5\textwidth}
\begin{algorithmic}
\MyFunction{$\tran{\varptr := \makeref(\varfld, \varint)}{\text{\memModel}}$}{$\absstat$}
\State $\cassign{\langle \scalarenv, \regfldseq, \addrseq, \absmemory\rangle}{\absstat}$
\State $\cassign{\scalarenv'}{\m{forget}(\scalarenv, \varptr)}$
\State $\Tlet \scalarenv'' = \m{addCons}($
\Statex \ind[6] $\scalarenv',  \varptr \neq 0)\Tin$
\State $\cassign{\addrseq'}{\m{forget}(\addrseq, \varptrbase)}$
\State $\langle \scalarenv'', \regfldseq, \addrseq', \absmemory\rangle$
\EndMyFunction
\State
\MyFunction{$\tran{\storeref(\varptr, \varfld, \varscalar)}{\text{\memModel}}$}{$\absstat$}
\State $\cassign{\langle \scalarenv, \regfldseq, \addrseq, \absmemory\rangle}{\absstat}$
\State $\cassign{\mb}{\mlocateabs(\varfld, \absmemory)}$
\State $\Tlet \langle \addrseq', \mb' \rangle \gets \invalidateCacheIfMiss^{\tdesignation}($
\Statex $\ind[6] \mb, \addrseq, \varptr)\Tin$
\State $\cassign{\langle \abscache, \abssummary, \langle \_, \_, \cacheinitpack \rangle \rangle}{\mb'}$
\State $\cassign{\abscache'}{\m{forget}(\abscache, \varfld)}$
\State $\cassign{\regfldseq'}{\m{forget}(\regfldseq, \varfld)}$
\State $\cassign{\regfldseq''}{\addEquality(\regfldseq', \varscalar, \varfld)}$
\State $\cassign{\absobjenv}{\langle\true, \true, \cacheinitpack\rangle}$
\State $\cassign{\mb''}{\langle \abscache', \abssummary, \absobjenv\rangle}$
\State $\langle \scalarenv, \regfldseq'', \addrseq',$
\Statex $\ind[7] \absmemory \setminus \{\mb\} \cup \{\mb''\}\rangle$
\EndMyFunction
\end{algorithmic}
\end{minipage}
\hspace{-.4em}
\vline
\hspace{-.8em}
\begin{minipage}[t]{0.56\textwidth}
\begin{algorithmic}
\MyFunction{$\tran{\varptrtwo, \varfldtwo := \gepref(\varptrone, \varfldone, \varint)}{\text{\memModel}}$}{$\absstat$}
\State $\cassign{\langle \scalarenv, \regfldseq, \addrseq, \absmemory\rangle}{\absstat}$
\State $\cassign{\scalarenv'}{\m{forget}(\scalarenv, \varptrtwo)}$
\State $\Tlet \scalarenv'' = \m{addCons}($
\Statex $\ind[3] \scalarenv, \varptrtwo = \varptrone + \varint)\Tin$
\State $\cassign{\addrseq'}{\m{forget}(\addrseq, \varptrtwobase)}$
\State $\Tlet \addrseq'' = \addEquality($
\Statex $\ind[6] \addrseq', \varptrtwobase, \varptronebase)\Tin$
\State $\langle \scalarenv'', \regfldseq, \addrseq'', \absmemory\rangle$
\EndMyFunction
\State
\MyFunction{$\tran{\varscalar := \loadref(\varptr, \varfld)}{\text{\memModel}}$}{$\absstat$}
\State $\cassign{\langle \scalarenv, \regfldseq, \addrseq, \absmemory\rangle}{\absstat}$
\State $\cassign{\mb}{\mlocateabs(\varfld, \absmemory)}$
\State $\Tlet \langle \addrseq', \mb' \rangle \gets \invalidateCacheIfMiss^{\tdesignation}($
\Statex $\ind[6] \mb, \addrseq, \varptr)\Tin$
\State $\cassign{\scalarenv'}{\m{forget}(\scalarenv, \varscalar)}$
\State $\cassign{\regfldseq'}{\m{forget}(\regfldseq, \varscalar)}$
\State $\cassign{\regfldseq''}{\addEquality(\regfldseq', \varfld, \varscalar)}$
\State $\langle \scalarenv', \regfldseq'', \addrseq',$
\Statex $\ind[7] \absmemory \setminus \{\mb\} \cup \{\mb'\}\rangle$
\EndMyFunction
\end{algorithmic}
\end{minipage}
\caption{Abstract transformers for memory operations.}
\vspace{-1em}
\label{figure:abs-transformer}
\end{figure}

\begin{figure}[t]
\begin{algorithmic}
\MyFunction{$\invalidateCacheIfMiss^{\tdesignation}$}{$\mb, \addrseq, \varptr$}
\State $\cassign{\langle\abscache, \abssummary, \langle\cacheused, \cachedirty, \cacheinitpack\rangle\rangle}{\mb}$
\Let{$\langle \addrseq', \mb'\rangle$}
    {\If{$\neg \cacheused \land \neg \testEquals(\addrseq, \varptrbase, \cachebase)$}
    \State $\cassign{\summary', \cacheinitpack'}{\mathbf{if}~ \cachedirty ~\mathbf{then}~ \lang{pack}^{\tdesignation}(\cache, \summary, \cacheinitpack) ~\mathbf{else}~ \summary, \cacheinitpack}$
    \State $\cassign{\cache'}{\lang{unpack}^{\tdesignation}({\summary}^{\prime})}$
    \State $\cassign{\addrseq'}{\m{forget}(\addrseq, \cachebase)}$
    \State $\cassign{\addrseq''}{\addEquality(\addrseq', \varptrbase, \cachebase)}$
    \State $\langle \addrseq'', \langle \cache', \summary', \langle\true, \false, \cacheinitpack'\rangle\rangle\rangle$
    \Else $~\langle\addrseq, \mb\rangle$
    \EndIf}
\EndLet $\langle \addrseq', \mb'\rangle$ 
\EndMyFunction
\MyFunction{$\lang{pack}^{\tdesignation}$}{$\abscache, \abssummary, \cacheinitpack$}{$~\mathbf{if}~ \neg\cacheinitpack ~\mathbf{then}~ \langle\lang{copy}(\abscache), \true\rangle ~\mathbf{else}~ \langle\abssummary \sqcup \abscache, \cacheinitpack\rangle $}
\EndMyFunction
\MyFunction{$\lang{unpack}^{\tdesignation}$}{$\abssummary$}{$~\lang{copy}(\abssummary)$}
\EndMyFunction
\end{algorithmic}
\caption{Abstract cache operations.}
\label{algo:cache}
\vspace{-1em}
\end{figure}

The most relevant transfer functions for inferring object invariants are shown in~\cref{figure:abs-transformer}. For the initial state of analysis, we assign all subdomain elements with $\top$, except for $\absobjenv$ in each memory bank as $\langle\false, \false,\false\rangle$. The third flag, $\cacheinitpack$, is $\false$ to indicate the $\abssummary$ does not represent any concrete objects.

For $\makeref$, the transformer assigns a $\varptr$ as not NULL in $\scalarenv$
indicating the valid address of the allocated object that $\varptr$ refers to.
For $\gepref$, the transformer computes the address for $\varptrtwo$ by addition
in $\scalarenv$ and establishes an equivalence between $\varptrtwo$ and
$\varptrone$ in $\addrseq$, denoting that the two pointers refer to the same memory object. 
For $\loadref / \storeref$, the transformer requires that the object referred by
$\varptr$ is in the cache before it is accessed. The function
$\invalidateCacheIfMiss^{\tdesignation}$ in~\cref{algo:cache} checks for a cache miss and handles operations when a miss happens. It tests whether $\varptr$ refers to the cached object by comparing
$\varptrbase$ with $\cachebase$ in $\addrseq$. When the cache is missed, the
function performs $\lang{pack}^{\tdesignation}$ and $\lang{unpack}^{\tdesignation}$.
The $\lang{pack}^{\tdesignation}$ operation merges $\abscache$ into
$\abssummary$.
The invariants of the first cached object are copied to $\abssummary$ because, initially, $\abssummary$ does not represent any concrete objects. We change the flag $\cacheinitpack$ to $\true$ since the $\abssummary$ now holds the invariants for that object. Any subsequent packs use the join operation.
The $\lang{unpack}^{\tdesignation}$ is achieved by copying the
$\abssummary$ as the new $\abscache$.
The $\lang{pack}^{\tdesignation}$ and $\lang{unpack}^{\tdesignation}$ operations are similar to the \emph{fold} and \emph{expand}
in~\cite{DBLP:conf/tacas/GopanDDRS04} but simpler because $\abscache$ and
$\abssummary$ are two domain values underlying the same field dimensions. 
After
unpacking, $\cachebase$ equals $\varptrbase$, signifying the $\abscache$ is
for the new MRU object. The transformer then performs a strong read/update in $\abscache$ without
changing any invariant stored in $\abssummary$. The read/update creates an
equivalence relation between $\varfld$ and $\varscalar$ in $\regfldseq$ through
$\addEquality$. For field read, the transformer discards the information in $\varscalar$
before adding the equality. For field update, the transformer forgets
information about $\varfld$ ahead of setting the equality and sets $\cachedirty$ to
$\true$ afterward.

Other abstract operators, including join, meet, widening, and narrowing, are
computed pointwise over subdomains with an additional caching step: packing the dirty cache
for each memory bank and resetting it as unused. 
The full definition for applying domain operators is available in the extended version of the paper \yusen{cite}.


We argue that the abstract semantics is sound as it is systematically derived
from the concrete semantics. At each program point, the scalar abstraction
over-approximates the set of numeric values or addresses of each scalar
variable. For memory objects, the abstraction collapses concrete objects in each
memory bank into one summary (abstract) object, also as an over-approximation.
The soundness argument follows from our design of abstraction and Galois connections.
We omit it here since the abstraction is straightforward.


\begin{figure}[t]

\scalebox{0.6}{
  \begin{tikzpicture}[
    basedomain/.style={rectangle, draw, fill=yellow!30, align=center, minimum width=1cm, minimum height=0.5cm, font=\scriptsize},
    eqdomain/.style={rectangle, draw, fill=purple!30, align=center, minimum width=1cm, minimum height=0.5cm, font=\scriptsize},
    booldomain/.style={rectangle, draw, fill=green!30, align=center, minimum width=1cm, minimum height=0.5cm, font=\scriptsize},
    composedomain/.style={rectangle, draw, fill=gray!20, rounded corners, align=center, minimum width=1cm, minimum height=1cm, font=\scriptsize},
    cartesian/.style={circle, draw, fill=blue!30, inner sep=0pt, minimum size=6mm},
    reduce/.style={circle, draw, fill=red!30, inner sep=0pt, minimum size=6mm},
    fields/.style={matrix of nodes, nodes in empty cells, ampersand replacement=\&},
    background/.style={rectangle, draw, fill=gray!10, rounded corners, inner sep=0.2cm},
    myarrow/.style={thick},
    reducearrow/.style={thick, fill=red}
]

\node(pic1) at (-5, -1) {
\begin{tikzpicture}
    \node (pc) [signal, minimum width=1.3cm, minimum height=0.5cm, align=center, draw=black, font=\small, above left=0.5cm and 0.7cm of odimap] {Loop entry};
    \node (odimap) [composedomain, minimum width=3.3cm, minimum height=3.6cm]  {};
    \node (basedom) [basedomain, minimum width=3.4cm, minimum height=1cm, above = 0.2cm of odimap] {$\mathit{sz} = 1 \land \mathit{i} = \mathit{sz} \land \ldots$};
    \node (flags) [booldomain, minimum width=2.05cm, minimum height=0.5cm, above right = 1cm and -.8cm of odimap] at (odimap.center) {};
    \node(flagname) [above right = -.1cm and -1.7cm of flags, font=\tiny] { \ttfamily{used dirty ispack}};
    \matrix(flagvalue)[fields, nodes={draw, anchor=center}, column sep=-0.7,row sep=.5, right = -.7cm of flags, font=\tiny] at (flags.center) { $\true$ \& $\true$ \& $\false$ \\};

    \node (cache) [basedomain, minimum width=2.5cm, minimum height=1.5cm, above = -0.6cm of odimap] at (odimap.center) {$\mathit{capacity} = 1  ~\land  $\\$ \mathit{length} = 0 ~\land$\\$\ldots$};
    \node (summary) [basedomain, minimum width=2.5cm, minimum height=1cm, below = 0.1cm of cache] {$\top$};
    \node (eqptrdom) [eqdomain, minimum width=2.5cm, minimum height=0.8cm, right = 0.2cm of basedom] {$p^\mathit{base} \approx \cachebase\land\ldots$};
    \node (eqfldscldom) [eqdomain, minimum width=2.5cm, minimum height=0.8cm, below = 0.2cm of eqptrdom] {$\mathit{sz} \approx \mathit{capacity}\land\ldots$};
    \node[below right] at (basedom.north west) {\small{$\scalarenv$}};
    \node[below right] at (odimap.north west) {\small{$\mb$}};
    \node[below right] at (eqfldscldom.north west) {\tiny{$\regfldseq$}};
    \node[below right] at (eqptrdom.north west) {\tiny{$\addrseq$}};
    \node[below right] at (cache.north west) {\tiny{$\cache$}};
    \node[below right] at (summary.north west) {\tiny{$\summary$}};
    \node[right = -.05cm of flags] at (flags.west) {\tiny{$\absobjenv$}};
 \end{tikzpicture}
};
\node(label1) [font=\large] at (-5.8, -3.9){(a) state at iteration~1: $s_1$};
\node(pic2) at (5, -1) {
\begin{tikzpicture}
    \node (odimap) [composedomain, minimum width=3.3cm, minimum height=3.6cm]  {};
    \node (pc) [signal, minimum width=1.3cm, minimum height=0.5cm, align=center, draw=black, font=\small, above left=0.5cm and 0.7cm of odimap] {Loop entry};
    \node (basedom) [basedomain, minimum width=3.4cm, minimum height=1cm, above = 0.2cm of odimap] {$\mathit{sz} = 2 \land \mathit{i} = \mathit{sz} \land \ldots$};
    \node (flags) [booldomain, minimum width=2.05cm, minimum height=0.5cm, above right = 1cm and -.8cm of odimap] at (odimap.center) {};
    \node(flagname) [above right = -.1cm and -1.7cm of flags, font=\tiny] { \ttfamily{used dirty ispack}};
    \matrix(flagvalue)[fields, nodes={draw, anchor=center}, column sep=-0.7,row sep=.5, right = -.7cm of flags, font=\tiny] at (flags.center) { $\true$ \& $\true$ \& $\true$ \\};

    \node (cache) [basedomain, minimum width=2.5cm, minimum height=1.2cm, above = -0.4cm of odimap] at (odimap.center) {$\mathit{capacity} = 2  ~\land  $\\$ \mathit{length} = 1 ~\land$\\$\ldots$};
    \node (summary) [basedomain, minimum width=2.5cm, minimum height=1.2cm, below = 0.1cm of cache] {$\mathit{capacity} = 1  ~\land  $\\$ \mathit{length} = 0 ~\land$\\$\ldots$};
    \node (eqptrdom) [eqdomain, minimum width=2.5cm, minimum height=0.8cm, right = 0.2cm of basedom] {$p^\mathit{base} \approx \cachebase\land\ldots$};
    \node (eqfldscldom) [eqdomain, minimum width=2.5cm, minimum height=0.8cm, below = 0.2cm of eqptrdom] {$\mathit{sz} \approx \mathit{capacity}\land\ldots$};
    \node[below right] at (basedom.north west) {\small{$\scalarenv$}};
    \node[below right] at (odimap.north west) {\small{$\mb$}};
    \node[below right] at (eqfldscldom.north west) {\tiny{$\regfldseq$}};
    \node[below right] at (eqptrdom.north west) {\tiny{$\addrseq$}};
    \node[below right] at (cache.north west) {\tiny{$\cache$}};
    \node[below right] at (summary.north west) {\tiny{$\summary$}};
    \node[right = -.05cm of flags] at (flags.west) {\tiny{$\absobjenv$}};
 \end{tikzpicture}
};

\node(label2) [font=\large] at (4.4, -3.9){(b) state after Line $11$: $s_b$};

\node(pic3) at (-5, -6.8) {
\begin{tikzpicture}
    \node (pc) [signal, minimum width=1.3cm, minimum height=0.5cm, align=center, draw=black, font=\small, above left=0.5cm and 0.7cm of odimap] {Loop entry};
    \node (odimap) [composedomain, minimum width=3.3cm, minimum height=3.6cm]  {};
    \node (basedom) [basedomain, minimum width=3.4cm, minimum height=1cm, above = 0.2cm of odimap] {$1\leq \mathit{sz} \leq 2 \land \mathit{i} = \mathit{sz} \land \ldots$};
    \node (flags) [booldomain, minimum width=2.05cm, minimum height=0.5cm, above right = 1cm and -.8cm of odimap] at (odimap.center) {};
    \node(flagname) [above right = -.1cm and -1.7cm of flags, font=\tiny] { \ttfamily{used dirty ispack}};
    \matrix(flagvalue)[fields, nodes={draw, anchor=center}, column sep=-0.7,row sep=.5, right = -.7cm of flags, font=\tiny] at (flags.center) { $\false$ \& $\false$ \& $\true$ \\};

    \node (cache) [basedomain, minimum width=2.5cm, minimum height=1cm, above = -0.1cm of odimap] at (odimap.center) {$\top$};
    \node (summary) [basedomain, minimum width=2.5cm, minimum height=1.5cm, below = 0.1cm of cache] {$1 \leq \mathit{capacity} \leq 2  ~\land  $\\$ \mathit{length} = \mathit{capacity} - 1 ~\land$\\$\ldots$};
    \node (eqptrdom) [eqdomain, minimum width=2.5cm, minimum height=0.8cm, right = 0.2cm of basedom] {$p^\mathit{base} \approx \cachebase\land\ldots$};
    \node (eqfldscldom) [eqdomain, minimum width=2.5cm, minimum height=0.8cm, below = 0.2cm of eqptrdom] {$\mathit{sz} \approx \mathit{capacity}\land\ldots$};
    \node[below right] at (basedom.north west) {\small{$\scalarenv$}};
    \node[below right] at (odimap.north west) {\small{$\mb$}};
    \node[below right] at (eqfldscldom.north west) {\tiny{$\regfldseq$}};
    \node[below right] at (eqptrdom.north west) {\tiny{$\addrseq$}};
    \node[below right] at (cache.north west) {\tiny{$\cache$}};
    \node[below right] at (summary.north west) {\tiny{$\summary$}};
    \node[right = -.05cm of flags] at (flags.west) {\tiny{$\absobjenv$}};
 \end{tikzpicture}
};
\node(label3) [font=\large] at (-5.8, -9.6){(c) state at iteration 2: $s_2 := s_1 \sqcup s_b$};

\node(pic4) at (5, -6.8) {
\begin{tikzpicture}
    \node (pc) [signal, minimum width=1.3cm, minimum height=0.5cm, align=center, draw=black, font=\small, above left=0.5cm and 0.7cm of odimap] {Loop entry};
    \node (odimap) [composedomain, minimum width=3.3cm, minimum height=3.6cm]  {};
    \node (basedom) [basedomain, minimum width=3.4cm, minimum height=1cm, above = 0.2cm of odimap] {$1\leq \mathit{sz} \leq +\infty \land \mathit{i} = \mathit{sz} \land \ldots$};
    \node (flags) [booldomain, minimum width=2.05cm, minimum height=0.5cm, above right = 1cm and -.8cm of odimap] at (odimap.center) {};
    \node(flagname) [above right = -.1cm and -1.7cm of flags, font=\tiny] { \ttfamily{used dirty ispack}};
    \matrix(flagvalue)[fields, nodes={draw, anchor=center}, column sep=-0.7,row sep=.5, right = -.7cm of flags, font=\tiny] at (flags.center) { $\false$ \& $\false$ \& $\true$ \\};

    \node (cache) [basedomain, minimum width=2.5cm, minimum height=1cm, above = -0.1cm of odimap] at (odimap.center) {$\top$};
    \node (summary) [basedomain, minimum width=2.5cm, minimum height=1.5cm, below = 0.1cm of cache] {$1 \leq \mathit{capacity} \leq +\infty  ~\land  $\\$ \mathit{length} = \mathit{capacity} - 1 ~\land$\\$\ldots$};
    \node (eqptrdom) [eqdomain, minimum width=2.5cm, minimum height=0.8cm, right = 0.2cm of basedom] {$p^\mathit{base} \approx \cachebase\land\ldots$};
    \node (eqfldscldom) [eqdomain, minimum width=2.5cm, minimum height=0.8cm, below = 0.2cm of eqptrdom] {$\mathit{sz} \approx \mathit{capacity}\land\ldots$};
    \node[below right] at (basedom.north west) {\small{$\scalarenv$}};
    \node[below right] at (odimap.north west) {\small{$\mb$}};
    \node[below right] at (eqfldscldom.north west) {\tiny{$\regfldseq$}};
    \node[below right] at (eqptrdom.north west) {\tiny{$\addrseq$}};
    \node[below right] at (cache.north west) {\tiny{$\cache$}};
    \node[below right] at (summary.north west) {\tiny{$\summary$}};
    \node[right = -.05cm of flags] at (flags.west) {\tiny{$\absobjenv$}};
 \end{tikzpicture}
};
\node(label4) [font=\large] at (4.4, -9.6){(d) state at fixpoint: $s_{\mathit{fix}} := s_1 \triangledown s_2$};

 \end{tikzpicture}}
\caption{Fixpoint computation for the entry state of the loop in~\cref{subfig:cfg}.}
\label{fig:trace}
\vspace{-1em}
\end{figure}
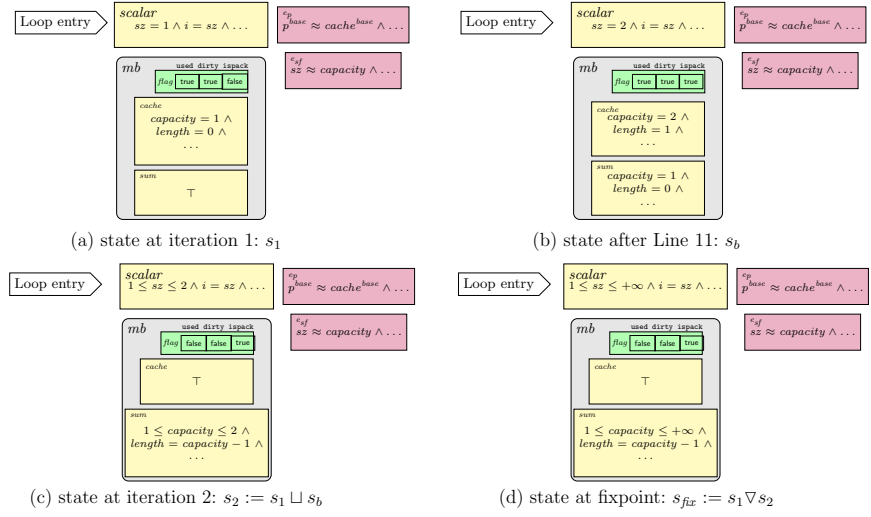

\cref{fig:trace} illustrates the computation of abstract states at the
loop entry of \cref{subfig:cfg}. In \cref{fig:trace}a, 
state $s_1$ represents the an abstract state at the loop entry after the first iteration opf the loop. 
Since during the first iteration only one \code{byte\_buf} object is initialized, 
the cache in $s_1$ has the invariants only of that object:
$\mathit{len} = 0$ and $\mathit{cap} = 1$, while the summary has no objects (i.e., $\cacheinitpack$ flag is unset). 
The next abstract state is
$s_b$ (\cref{fig:trace}b) after \cref{irline:final}. During the second iteration, the
cache is flushed for the new \code{byte\_buf} object and the summary only maintains the invariants
for the flushed object. Then, $s_1$ and $s_b$ are joined at the loop entry, resulting in $s_2$ (\cref{fig:trace}c). The join is pairwise across subdomains after
the caches of both states are flushed. Finally, the widening operator is applied
to reach a fixpoint, as shown in \cref{fig:trace}d.

\begin{figure}[t]
\begin{algorithmic}
\MyFunction{$\m{reduce}$}{$\mathit{base}_{\mathit{src}}, \mathit{base}_{\mathit{dst}}, \mathit{e}$}
\State $\cassigntwo{e'}{\m{project}(e, \mathcal V_{\mathit{src}} \cup \mathcal V_{\mathit{dst}})}{cons}{\m{toCons}(\mathit{e'})}$
\State $\cassign{\mathit{base}^{'}_{\mathit{dst}}}{\m{project} ( (\mathit{base}_{\mathit{dst}} \sqcap \m{addCons}(\mathit{base}_{\mathit{src}}, cons)), \mathcal V_{\mathit{dst}})}$
\State $\mathit{base}^{'}_{\mathit{dst}}$
\EndMyFunction
\State
\MyFunction{$\m{reduction}$}{$\absstat$}
\State $\cassign{\langle \scalarenv, \regfldseq, \addrseq, \absmemory\rangle}{\absstat}$
\ForAll{$\mb \in \absmemory$} \Comment{Step 1: reduce from caches to base}
  \State $\cassign{\langle \abscache, \_, \_ \rangle}{\mb}$
  \State $\scalarenv' := \m{reduce}( \abscache, \scalarenv, \regfldseq)$
\EndFor
\ForAll{$\mb \in \absmemory$} \Comment{Step 2: reduce from base to caches}
  \State $\cassign{\langle \abscache, \abssummary, \absobjenv \rangle}{\mb}$
  \State $\cassign{\abscache'}{\m{reduce}(\scalarenv', \abscache, \regfldseq)}$
  \State $\mb := \langle \abscache', \abssummary, \absobjenv \rangle$ \Comment{Update $\mb$ directly}
\EndFor
\State $\langle \scalarenv', \regfldseq, \addrseq, \absmemory\rangle$
\EndMyFunction
\end{algorithmic}
\caption{Domain reduction.}
\label{algo:object-reduction}
\vspace{-1em}
\end{figure}
As the memory and scalar properties are kept separately, we configure a domain
reduction step to exchange information between each bank's $\abscache$ and
$\scalarenv$ through the equalities that are introduced during $\loadref$ and
$\storeref$. We use a bidirectional reduction (see red arrows on the right
of~\cref{fig:domain_hierarchy}): one direction flows from the $\AbsCache$ of
each memory bank to $\Basedom$; the other is in the opposite. The domain
reduction follows~\cref{algo:object-reduction} which reduces an abstract state
as $\absstat$ in two steps by propagates numerical properties (1) from each
$\abscache$ into $\scalarenv$, and (2) from the $\scalarenv$ back to each
$\abscache$. The algorithm computes the iterated pairwise reduction through
$\m{reduce}$ which operates on each bank's $\abscache$ and $\scalarenv$. For
example, \cref{fig:absstate-emp} shows the $\abscache$ after applying the
reduction whose values are refined for \code{cap} and \code{len} based on
equalities generated for field updates through scalars \code{sz} and \code{i} in
$\scalarenv$. The $\abscache$ is reduced through the step (2) which involves
$\m{reduce}$ converting equalities ($\mathit{len}\approx i$ and
$\mathit{cap}\approx \mathit{sz}$) into linear constraints and adding them to
$\scalarenv$. Then, it performs a meet with $\abscache$ to propagate numerical
information from $\scalarenv$. Finally, it projects the result of the meet
to the field variables, and obtains the new $\abscache$.

When $\m{reduction}$ is executed once, it refines the abstract values in each
bank's $\abscache$ and $\scalarenv$ in the state. It adds numerical properties
and preserves equalities. This ensures that it is both reductive and sound. We
terminate the reduction after one iteration for each of the two directions.

In summary, we introduce \mrudomain, a composite abstract domain and its corresponding transformer for inferring object invariants. As a reduced product of domains for scalars and objects, \mrudomain is effective for scalable analysis. The reduction algorithm leverages equalities between variables to avoid precision loss.
\section{Implementation}
\label{sec:implementation}
We have implemented the \mrudomain\footnote{Publicly available at \texttt{https://github.com/LinerSu/crab/tree/VMCAI-2024}.} in
\crab~\cite{DBLP:conf/vstte/GurfinkelN21}, a library for building abstract
interpretation-based analyses. The $\AbsMemory$ is implemented using a Patricia
tree~\cite{okasaki1998fast} for \emph{structural sharing} among multiple
abstract elements during analysis. This approach prevents redundant copying of
domain values when computing the outputs of domain operators and transfer functions, 
allowing efficient memory sharing for parts of the abstract state that remain unchanged after an operation.
For example, two domain elements of
$\AbsMemory$ share memory banks if they are unchanged during computation.

We have developed a custom equality domain based on a union-find data structure
to represent variable equivalence (e.g., $x \approx y$). The details of this
domain are available the extended version of the paper \yusen{cite}. Each equivalence class corresponds to
a set of variables (e.g., $\{p^\mathit{base}, \cachebase\}$ as $p^\mathit{base}
\approx \cachebase$ in \cref{fig:absstate-emp}). This structure fits the representation of equivalence relations and efficiently supports domain operation.
Our implementation also partitions $\eqsfdom$ into reduced product of smaller
domains for better alignment with variable
packing~\cite{DBLP:conf/pldi/BlanchetCCFMMMR03}. Specifically, we use an
equality domain $\eqregsdom$ for scalars and $\eqfldsdom$, in each memory bank,
for fields. The domain value of $\eqsfdom$ is the union of these smaller domain
values. For example, $i\approx \mathit{len} \land \mathit{sz} \approx
\mathit{cap}$ is maintained as two classes $\regfldseq := \{i, \mathit{len}\},
\{\mathit{sz}, \mathit{cap}\}$ which are equivalent to splitted classes as
$\regseq := \{i, \tilde{a}\}, \{sz, \tilde{b}\}$ and $\fldseq:=\{\mathit{len},
\tilde{a}\}, \{\mathit{cap}, \tilde{b}\}$ with special representatives
$\tilde{a}, \tilde{b}$.

For memory partitioning, we use \seadsa~\cite{DBLP:conf/sas/GangeNSSS16} to divide the memory used by the program into memory banks, with each bank containing objects from the
same allocation site. As mentioned earlier, in \crabir, a field variable
represents an offset to access an object field. The $\mlocate$ function of
\memModel is defined by mapping fields to their corresponding bank. However, in
practice, not all field offsets can be determined statically. We
over-approximate the values of such field by $\top$. Improving this is left for
future work.

For effective and efficient domain reduction, we use heuristics to balance
precision and performance. \mrudomain tracks which direction needs reduction.
For example, if equalities between fields and scalars only affect memory reads,
there is no need to apply a reduction to refine the corresponding cache.
We also allow reduction to be performed on demand.
For instance, reduction is applied when an assertion is present in the program.

\section{Evaluation}
\label{sec:experiments}
We performed three kinds of experiments: {\bf scale}, {\bf precision}, and {\bf case study}. All experiments were conducted on a desktop computer with an Intel Xeon E$5$-$2680$ @$2.50$GHz, with $256$ GB RAM, and are available at \url{https://doi.org/10.5281/zenodo.13849174}.

\begin{figure}[t]
  \centering
  \includegraphics[width=.7\linewidth]{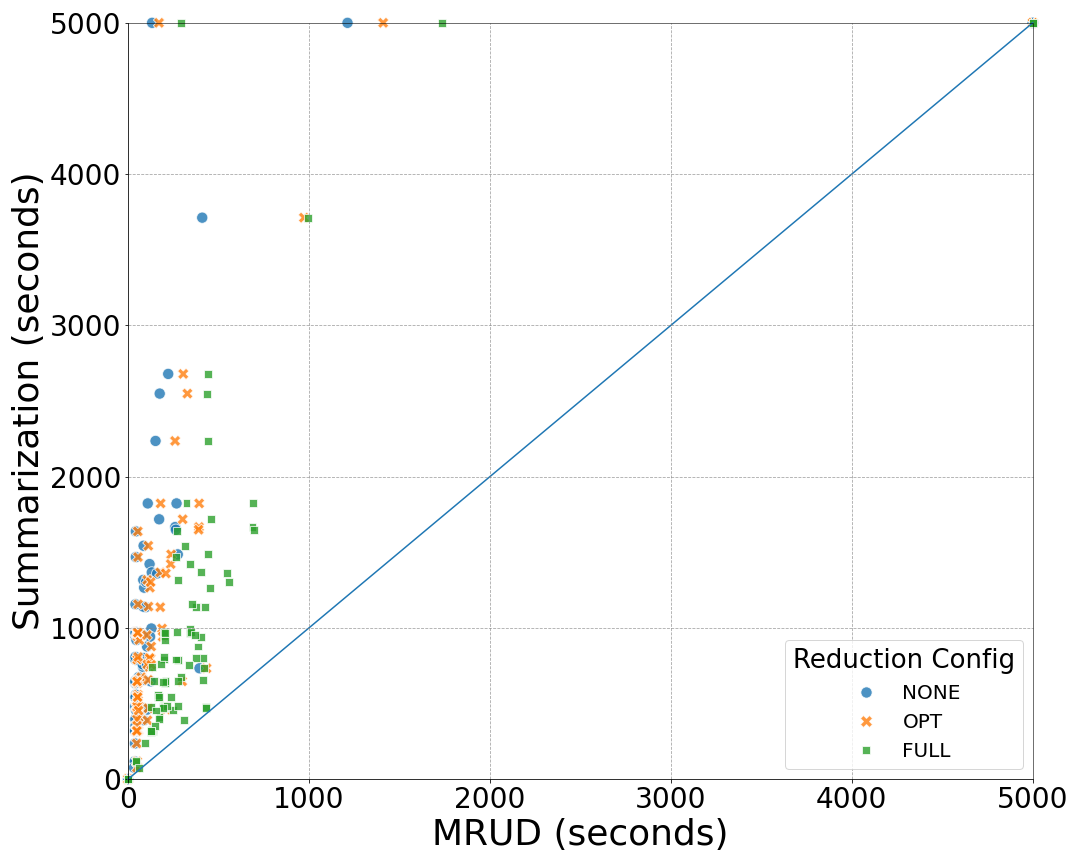}
  \captionof{figure}{Scalability results. Summarization refers to $\mathcal{D_{S}}$ and \mrudomain to $\mathcal{D_{O}}$.}
  \label{fig:scale}
  \vspace{-1em}
\end{figure}

First, the {\bf scale} experiment compares the performance of \mrudomain ($\mathcal{D_{O}}$) with the summarization-based~\cite{DBLP:conf/vstte/GurfinkelN21} domain ($\mathcal{D_{S}}$) from \crab by timing analysis of $114$ programs: $5$ from~\cite{DBLP:conf/vstte/GurfinkelN21}, and $109$ from GNU Coreutils~\cite{gnuCoreutilsWeb}. We used the Zones\footnote{The zones domain represents all the binary relationships between two-variable difference (including zero), stored in a Difference-Bound Matrix (DBM).}~\cite{DBLP:conf/sas/GangeNSSS16} abstract domain for its simplicity and sufficiency in expressing (relational) memory safety invariants. The primary goal is to show that $\mathcal{D_{O}}$ scales better than $\mathcal{D_{S}}$ due to the effect of variable packing~\cite{DBLP:conf/pldi/BlanchetCCFMMMR03} in $\mathcal{D_{O}}$ that follows from representing each partition with a different DBM, while $\mathcal{D_{S}}$ relies on a single DBM for expressing all scalars (included ghost ones) and summary variables. Another goal is to measure the overhead introduced by domain reduction, which incurs extra costs.
To evaluate this, we provide two additional strategies: FULL, which applies reduction at each transfer function, and NONE, where no reduction is applied, and compare them with the heuristic strategy, OPT. These three strategies highlight the different costs of reduction.

\cref{fig:scale} shows the timing results, with a timeout of $5\,000$ seconds per program. Both domains time out on $6$ cases, while $\mathcal{D_{S}}$ times out on $2$ more cases. Excluding timeout cases, $\mathcal{D_{O}}$ outperforms $\mathcal{D_{S}}$ on nearly every benchmark. On average, $\mathcal{D_{O}}$ with NONE, OPT, and FULL configurations is $81$x, $76$x, and $57$x faster than $\mathcal{D_{S}}$, respectively. This demonstrates the advantage of composite abstract domains for inferring object invariants in large and complex programs, regardless of the domain reduction strategy used.

We analyze \code{ginstall} from GNU Coreutils to understand why $\mathcal{D_{O}}$ is faster. The running time for $\mathcal{D_{S}}$ is  $1\,846$s, while for $\mathcal{D_{O}}$, it takes $273$s. Most of the time in both domains is spent on join operations, where $\mathcal{D_{S}}$ spends $600$s, while $\mathcal{D_{O}}$ takes $95$s. 
Joining in $\mathcal{D_{O}}$ is also efficient because it allows to share DBMs across memory banks from other states (structural sharing for $\AbsMemory$ domain). Another reason is that most DBMs in $\mathcal{D_{O}}$ are small, making their joins less costly compared to $\mathcal{D_{S}}$, where large DBMs are involved. This efficiency is also reflected in the time to copy DBMs: $\mathcal{D_{S}}$ takes $260$s, while $\mathcal{D_{O}}$ takes $20$s.

As for domain reduction, applying it at each transfer function is inefficient, as FULL takes $144$ ($177$) seconds longer than OPT (NONE) on average. The heuristics strategy (OPT) effectively handles complex programs without significant performance loss.

\begin{table}[t]
\centering
\begin{minipage}[b]{0.4\linewidth}
\lstset{escapeinside={(*@}{@*)}}
\begin{lstlisting}[basicstyle=\footnotesize, language=C, style=mystyle]
void foo(){
  char ary1[1], ary2[2];
  struct byte_buf o1 = {.len = 0, .cap = 1, .buf=ary1};
  struct byte_buf o2 = {.len = 1, .cap = 2, .buf=ary2};
  struct byte_buf *p;
  if (/*some conditions*/) {
    p = &o1;
  } else {
    p = &o2;
  }
  p->len = 15; p->cap = 20;(*@\label[cline]{cline:emp2}@*)
  ...
}
\end{lstlisting}
\captionof{figure}{Another C program.}
\label{fig:alias-example}
\vspace{-1em}
\end{minipage}
\hspace{0.3cm} 
\begin{minipage}[b]{0.5\linewidth}
  \centering
\scalebox{0.9}{
\begin{tabular}{lcp{0.001\textwidth}cp{0.001\textwidth}ccp{0.001\textwidth}cc}
\multicolumn{1}{c}{\multirow{2}{*}{Program}} & \multirow{2}{*}{\#A} &  & $\mathcal{D_{O}}$ &  & \multicolumn{2}{c}{$\mathcal{D_{S}}$} &  & \multicolumn{2}{c}{$\mathcal{D_{R}}$} \\ \cline{4-4} \cline{6-7} \cline{9-10} 
\multicolumn{1}{c}{}                         &                      &  & safe   &  & safe         & warn        &  & safe        & warn        \\ \cline{1-2} \cline{4-4} \cline{6-7} \cline{9-10} 
bytebuf                                      & 3                    &  & 3      &  & 0            & 3           &  & 0           & 3           \\
bytebuf\_memcpy                              & 3                    &  & 3      &  & 0            & 3           &  & 0           & 3           \\
bytebuf\_path                                & 3                    &  & 3      &  & 1            & 2           &  & 1           & 2           \\
ipc\_handler                                 & 3                    &  & 3      &  & 2            & 1           &  & 2           & 1           \\
mult\_bytebuf                                & 3                    &  & 3      &  & 0            & 3           &  & 0           & 3           \\
object                                       & 1                    &  & 1      &  & 0            & 1           &  & 0           & 1           \\
range                                        & 2                    &  & 2      &  & 1            & 1           &  & 0           & 2           \\ \hline
\end{tabular}}
\vspace{0.3cm}
\caption{Precision results.}
\label{table:precision}
\vspace{-1em}
\end{minipage}
\end{table}

Second, the {\bf precision} experiment compares $\mathcal{D_{O}}$ against existing heap abstract domains: $\mathcal{D_{S}}$ and Mopsa with recency abstraction ($\mathcal{D_{R}}$). Since all three domains follow allocation-site abstraction, which summarizes multiple objects into one and treats them indistinguishable, it becomes challenging to precisely track field updates on individual concrete objects. Specifically, $\mathcal{D_{S}}$ cannot overcome this limitation. $\mathcal{D_{R}}$ improves precision by differentiating the most recently allocated object at the same site. $\mathcal{D_{O}}$ provides a more general strategy by distinguishing the most recently used object. As a result, $\mathcal{D_{O}}$ still precisely models field updates after object initialization, such as field updates on \cref{cline:length,cline:ubuffer} in \cref{fig:discuss-example}, which either $\mathcal{D_{R}}$ or $\mathcal{D_{S}}$ cannot handle.

Another challenge is path sensitivity since unclear pointer aliasing leads to imprecise modeling of field updates. For example, in \cref{fig:alias-example}, two \code{byte\_buf} objects, \code{o1} and \code{o2}, are allocated separately, and a pointer \code{p} is referred to either \code{o1} or \code{o2}. Modeling strong field updates in \cref{cline:emp2} requires knowing which object is being updated, but it is unknown which object the pointer \code{p} refers to. Both $\mathcal{D_{R}}$ and $\mathcal{D_{S}}$ can track field updates precisely, but they need more accurate points-to information. $\mathcal{D_{O}}$, however, allows strong updates by placing \code{o1} and \code{o2} in the same memory bank. When updating a field on either object, we load it into the cache and perform strong updates without precise pointer aliasing.

We provide a set of $7$ benchmarks\footnote{Available at: \texttt{https://github.com/LinerSu/MRU-Domain-Benchmarks}.} with similar code pattern like examples in \cref{fig:discuss-example,fig:alias-example} for evaluation and configure all three domains using the octagon domain. \cref{table:precision} shows that $\mathcal{D_{O}}$ successfully proves all assertions, showing the effectiveness of our methodology in providing a more precise memory abstraction. Conversely, $\mathcal{D_{S}}$ and $\mathcal{D_{R}}$ largely fail due to weak updates, as discussed above.

\newcommand{\aibmc}{AI4BMC\xspace}

Third, we present a {\bf case study} which integrates an Abstract Interpreter (AbsInt) into a Bounded Model Checker (BMC) pipeline for memory safety verification. This new pipeline, \aibmc, uses AbsInt to verify and remove a number of assertions before passing the problem to the SMT solver.

\begin{figure}[t]
  \centering
  \includegraphics[width=\linewidth]{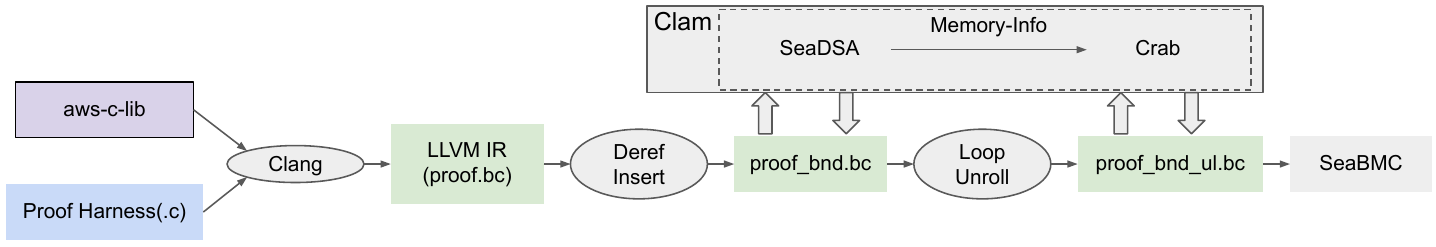}
  \caption{The \aibmc pipeline.}
  \label{fig:pipeline}
\end{figure}

The \aibmc pipeline, shown in~\cref{fig:pipeline}, starts by compiling and instrumenting the input program with buffer overflow checks. Next, AbsInt is applied to remove as many of these checks as possible. Now, the program still keeps the original loops. Then, the loops are unrolled using a user-supplied bound for BMC. Later, we run another AbsInt round to eliminate buffer overflow checks in the simplified program with unrolled loops. Last, we continue with the BMC pipeline, as in~\seabmc~\cite{DBLP:conf/fmcad/PriyaSBZVG22}, that generates a Verification Condition (VC) in SMT-LIB and uses an SMT-solver to check the VC's satisfiability such that the original program is safe if and only if SMT-LIB formula is unsatisfiable.

The motivation for \aibmc is that many memory safety arguments are simple and are established independently of loop bounds. We expect AbsInt to verify those, leaving less work for BMC. Thus, we consider \aibmc pipeline successful if (a) AbsInt discharges some buffer overflow checks before loop unwinding, and (b) \aibmc requires less overall runtime than the BMC pipeline.

We developed two benchmark suites from industrial code. The first is based on \texttt{aws-c-commons} verification tasks, where we reduce assertions only to memory safety. The second is based on a more complex code from AWS C SDK in C99 implementation. Together, there are $109$ verification tasks. The benchmarks\footnote{Available at \texttt{https://github.com/LinerSu/verify-c-common/tree/VMCAI-2025}.} have been adapted to simplify control flow since proving all memory safety checks requires path-sensitivity.

\begin{figure}[t]
\centering
  \includegraphics[width=.7\linewidth]{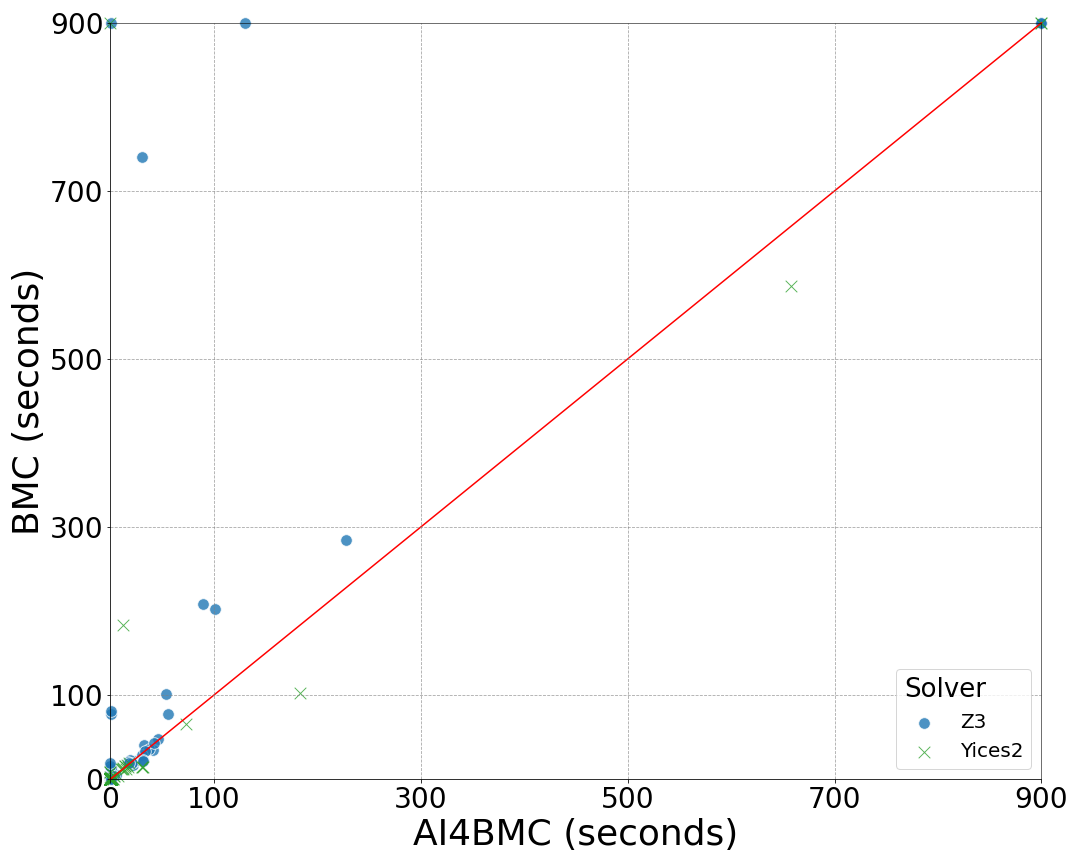}
  \caption{\aibmc vs. BMC.}
  \label{fig:ai4bmc}
\end{figure}

We evaluate the effectiveness \aibmc by comparing it with \seabmc which was previously compared against other state-of-the-art tools in~\cite{DBLP:conf/fmcad/PriyaSBZVG22}. Our performance evaluation focuses on these metrics: 
\begin{inparaenum}[(1)] 
  \item \textit{Faster} indicates \aibmc outperforms BMC;
  \item \textit{Slower} means \aibmc is slower than BMC; 
  \item \textit{AbsInt Time} expresses the run-time of AbsInt in the \aibmc pipeline.
\end{inparaenum}
For precision, we provide the \textit{AbsInt Solving Rate}, showing how many checks are solved before or after loop unrolling (LU). 
We used \mrudomain for \crab (AbsInt) and chose two SMT-solvers for \seabmc:  \zthree\footnote{We fixed the performance issue on Z3. The one we used is available at: \texttt{https://github.com/LinerSu/z3/tree/fix-performance}.}~\cite{DBLP:conf/tacas/MouraB08}, and \ytwo~\cite{DBLP:conf/cav/Dutertre14}. Experiments were conducted under $900$ seconds timeout and all results are summarized in~\cref{fig:ai4bmc} and~\cref{table:casestudy}.

\begin{table}[t]
\scalebox{0.8}{
\begin{tabular}{cp{0.001\textwidth}cp{0.001\textwidth}cp{0.001\textwidth}cp{0.001\textwidth}c}
\multirow{2}{*}{\textbf{Category}}      &  & \multirow{2}{*}{\textbf{Metric}}                       &  & \multirow{2}{*}{\textbf{\% Metric}}                          &  & \multicolumn{3}{c}{\textbf{Number of Cases}} \\ \cline{7-9} 
                                        &  &                                               &  &                                                     &  & \aibmc(Z3)         &          & \aibmc(Y2)         \\ \cline{1-1} \cline{3-3} \cline{5-5} \cline{7-7} \cline{9-9} 
\multirow{4}{*}{Performance Comparison} &  & \multirow{2}{*}{\textit{Faster} (Time Difference $>5$s)}                       &  & $> 95\%$ &  &     $10$        &          &     $2$       \\
                                        &  &                                               &  & others            &  &   $6$          &          &  $1$          \\ \cline{3-3} \cline{5-5} \cline{7-7} \cline{9-9} 
                                        &  & \multirow{2}{*}{\textit{Slower} (Time Difference $>5$s)}                       &  & $\leq 50\%$     &  &    $4$         &          &    $2$        \\
                                        &  &                                               &  & others              &  & $0$            &          &  $4$          \\ \cline{1-1} \cline{3-3} \cline{5-5} \cline{7-7} \cline{9-9} 
\multirow{2}{*}{AbsInt Performance}     &  & \multirow{2}{*}{\textit{AbsInt Time} in AI4BMC time}        &  & $> 40\%$                  &  &  $65$           &          &  $74$          \\
                                        &  &                                               &  & $\leq 40\%$                  &  &  $39$           &          &    $29$        \\ \cline{1-1} \cline{3-3} \cline{5-5} \cline{7-7} \cline{9-9} 
\multirow{4}{*}{Precision}              &  & \multirow{2}{*}{\textit{AbsInt Solving Rate} before LU} &  & $100\%$                           &  &    $37$         &          &     $37$       \\
                                        &  &                                               &  & $> 50\%$               &  & $52$            &          &    $52$        \\ \cline{3-3} \cline{5-5} \cline{7-7} \cline{9-9} 
                                        &  & \multirow{2}{*}{\textit{AbsInt Solving Rate} after LU}  &  & $100\%$                          &  &   $6$          &          &     $6$       \\
                                        &  &                                               &  & $> 50\%$              &  &  $1$           &          &  $1$          \\ \cline{1-1} \cline{3-3} \cline{5-5} \cline{7-7} \cline{9-9} 
\end{tabular}}
\caption{\aibmc vs. BMC details.}
\label{table:casestudy}
\end{table}

First, comparing performance between \aibmc and BMC. With \zthree, \aibmc timed out in $5$ cases, while BMC timed out in $7$ cases; AbsInt helped solving $2$ more cases. Excluding timeouts, \aibmc is at least 5s \emph{faster} than BMC in $16$ cases. The speed-up comes from AbsInt proving and discharging assertions checks. In $10$ of these $16$ cases, the speed-up exceeds over $95\%$, with AbsInt completely solving the checks in $9$ cases. The other $6$ cases show at least a $20\%$ speed-up. AbsInt takes under one second on average in all 16 cases.
There are $4$ cases in which \aibmc is at least 5s \emph{slower} than BMC. In two of these, the slowdowns are due to \zthree taking $6$s extra solving time on average, which is not surprising since the SMT performance is not always deterministic. In the other two, although \zthree solving time is decreased, AbsInt slows down by taking around $11$s, roughly a third of the total run-time.

The results with \ytwo are similar, but \ytwo is faster and exhibits better stability. Both \aibmc and BMC timed out in $4$ cases and $5$ cases individually, with $1$ case where AbsInt improves performance. \aibmc outperforms BMC in $3$ cases with at least a $93\%$ improvement. However, \aibmc is slower in $6$ cases, $4$ of which are affected by the slowdown of AbsInt. The other $2$ cases are due to the slows down of \seabmc and \ytwo. The \seabmc experiences a slowdown due to lambda-encoding, where the beta-reduction simplification time is not deterministic. While switching to array-encoding shows the effectiveness of AbsInt, this slows overall performance for both \aibmc and BMC.

Overall, the performance results show that AbsInt improves the overall performance of using BMC regardless of the solver used.

Second, in evaluating the performance of AbsInt, runtime ratios depend on the total running time of \aibmc and the solver selected. With \zthree, AbsInt takes over $40\%$ of the time on $65$ cases, but these cases terminate within $50$s, with AbsInt averaging only $0.1$s and maxing at $1.2$s. For the rest of the $39$ cases, AbsInt takes $40\%$ or less, with $5$ cases exceeding $50$s and $34$ cases under $50$s. For these $5$ longer cases, AbsInt accounts for under $2\%$, averaging $1$s with a maximum of $1.5$s. For the $34$ shorter cases, AbsInt contribution was below $36\%$. With \ytwo, the runtime percentage of AbsInt increases because \ytwo is efficient, with more cases where AbsInt accounts for a significant portion of the runtime. In summary, using AbsInt has no big cost, compared with the solving time of SMT solver. 

Last, for assertion rate, AbsInt solved more than $50\%$ of assertions in $89$ cases before LU, completely solving $37$ cases, and in $7$ cases after LU, fully solving $6$ cases. We only have $8$ cases where AbsInt solves less than half of the checks. The reasons are:
(1) the widening operation produces too imprecise invariants that cannot be recovered by narrowing. AbsInt needs more precise widening techniques to prove more checks; (2) Some memory safety invariants cannot be expressed by Zones or Octagons, and instead require more complex abstract domains such as Polyhedra; (3) Memory safety checks for C string require tracking the length of strings that our implementation does not support. We believe using~\cite{DBLP:conf/sas/JournaultMO18} to determine the null character of each string will improve overall precision. 

In this case study, we demonstrate the effectiveness of using AbsInt in the BMC pipeline. By using the Zones, it proves most memory safety checks in this industry project and reduces the number of checks BMC handles. This speeds up both BMC encoding and SMT solver performance.

\section{Related Works}

To deal with a potentially unbounded number of memory objects, most abstract analysis frameworks group memory objects together into \emph{summary objects} (e.g., \cite{DBLP:conf/tacas/GopanDDRS04}). A summary object represents properties that are common to all objects it stands for. The most common summarization is \emph{Allocation Site Abstraction} (ASA)~\cite{DBLP:conf/pldi/ChaseWZ90} that groups objects by their allocation site. In ASA, all concrete objects allocated at a certain line of a program are represented by one abstract summary object. Since each summary object represents a set of objects, it supports only \emph{weak} updates -- an assignment to the field of an object does not override previous value, but rather adds to it, to capture that the field update may modify only one object out of the summary. This significantly degrades analysis precision. 

The loss of precision is specifically important during object creation, when an object is first allocated and then initialized field-by-field. In ASA, because of weak updates, this results in all properties of the summary being lost since the newly allocated object has no properties in common with already summarized objects. A common solution, e.g., used by Mopsa, is \emph{recency abstraction}~\cite{DBLP:conf/sas/BalakrishnanR06} that refines ASA into: (a) the most recently allocated object, and (b) the rest. Since most recent object is a singleton, it can be updated \emph{strongly}, i.e., field updates overwrite previous values. Our approach is a further refinement that separates objects not by recency of \emph{creation}, but by recency of \emph{use}. 
In principle, other extensions of recency, such as~\cite{DBLP:conf/sas/BalatsourasS16} can be combined with our technique for further precision improvement.

The temporary isolation of recenctly-used objects avoids invariant violations in summarized objects during individual field updates. Our pack and unpack methods communicate changes between these two types of objects. This is similar to corresponding methods in~\cite{DBLP:journals/entcs/ChangRL05}, where the annotated \emph{pack/unpack} statements manage transitions of mutable objects during class method calls, allowing temporary updates while maintaining class invariants (i.e., invariants for all instances of a given class). Similarly, JayHorn~\cite{DBLP:conf/lpar/KahsaiKRS17} uses \emph{push/pull} statements for encoding each memory access. Each \emph{pull} statement reads fields of an object to make invariants available, while a following \emph{push} statement updates fields to ensure modifications preserve invariants. The concept of \emph{pack/unpack} has been used in refinement types~\cite{DBLP:conf/popl/RondonKJ10}, where the inference algorithm obtains predicates with \emph{fold/unfold} operations to prevent temporary invariant violations of objects from the same allocation site. Unlike our work, all prior work uses heuristics to manage placement of \emph{fold/unfold} operations. In contrast, our analysis automatically processes these during analysis.

The domain hierarchy in our \mrudomain uses two strategies. First, variable packing~\cite{DBLP:conf/pldi/BlanchetCCFMMMR03} is used to pack program variables for fields of memory objects in each memory bank. With two numerical domains per pack, our approach allows for the independent updating of invariants for each bank. The packing is rarely used in computing memory properties, but Toubhans et al.~\cite{DBLP:conf/sas/ToubhansCR14} introduced a product of memory domains that pack variables used for lists, trees, and other fixed-size structures. 
Second, domain reduction~\cite{DBLP:conf/popl/CousotC79} helps exchange equivalences between scalars and object fields. This is commonly used when abstract domains are organized modularly. Astr\'ee~\cite{DBLP:conf/asian/CousotCFMMMR06} combines various abstract domains in a sequence, using reduction steps for forward and backward propagation of information between them. \cite{DBLP:conf/fossacs/CousotCM11} interprets the Nelson-Oppen procedure as a domain reduction, propagating (dis)equalities across domains.

\section{Conclusion}
In this work, we present a new methodology for inferring object invariants that avoids temporarily breaking invariants following the concept of caching. Our new abstract domain, parameterized by numerical and equality domains, organizes a structured hierarchy, enabling scalable analysis of complex programs. We design a reduction algorithm following equalities introduced across numerical domains to avoid significant precision loss. Our results demonstrate that \mrudomain enhances both precision and scalability and can be effectively integrated with other verification techniques for memory safety.

%
%
%

%

\clearpage
\appendix

\section{Domain Operation}
\label{sec:domop}

We present domain lattice operations for the \mrudomain. The generic algorithm for join, meet, widening, and narrowing operations is shown in \cref{fig:latticeop}. The algorithm invokes a helper function in \cref{fig:flushcache} to clear the cache in each memory bank. After this, the subdomains are ready for pairwise operations. 

\begin{figure}[t]
\begin{algorithmic}
\MyFunction{$op^{\tdesignation}$}{$\absstat_1, \absstat_2$}
\State $\cassign{\langle \scalarenv_1, \regfldseq_1, \addrseq_1, \absmemory_1\rangle}{\absstat_1}$
\State $\cassign{\langle \scalarenv_2, \regfldseq_2, \addrseq_2, \absmemory_2\rangle}{\absstat_2}$
\ForAll{$\mb_1 \in \absmemory_1$}
  \State $\mb_1 := \m{flushCache}^{\tdesignation}(\mb_1)$ \Comment{Update $\mb_1$ directly}
\EndFor
\ForAll{$\mb_2 \in \absmemory_2$}
  \State $\mb_2 := \m{flushCache}^{\tdesignation}(\mb_2)$ \Comment{Update $\mb_2$ directly}
\EndFor
\State $\langle \scalarenv_1 ~op^{\tdesignation}~ \scalarenv_2, \regfldseq_1 ~op^{\tdesignation}~ \regfldseq_2, \addrseq_1 ~op^{\tdesignation}~ \addrseq_2, \absmemory_1 ~op^{\tdesignation}~ \absmemory_2 \rangle$
\EndMyFunction
\end{algorithmic}
\caption{Generic algorithm for the lattice operations.}
\label{fig:latticeop}
\end{figure}

\begin{figure}[t]
\begin{algorithmic}
\MyFunction{$\m{flushCache}^{\tdesignation}$}{$\mb$}
\If{$\neg \cacheused$}
    \State $\cassign{\summary', \cacheinitpack'}{\mathbf{if}~ \cachedirty ~\mathbf{then}~ \lang{pack}^{\tdesignation}(\cache, \summary, \cacheinitpack) ~\mathbf{else}~ \summary, \cacheinitpack}$
    \State $\langle \cache, \summary', \langle\false, \false, \cacheinitpack'\rangle\rangle$
\Else $~\mb$
\EndIf
\EndMyFunction
\end{algorithmic}
\caption{Helper function for the cache flush.}
\label{fig:flushcache}
\end{figure}

\section{Lightweight Equality Abstract Domain}
\label{sec:eqdom}
\newcommand{\vV}{\mathcal{V}}

In this section, we propose a lightweight equality domain, \neweqdomain, for representing variable equivalence of the form $x \approx y$. The section describes certain definitions, lower-level representation, and operations of domain.
\subsection{Representing Equality} A natural way to capture equality is with a union-find data structure: each equivalence class in a union-find corresponds to a set of equivalent variables. Formally, let $\vV$ be a finite set of variables, equivalence classes are represented as a partition $\vV_1, \ldots, \vV_n$ of $\vV$. To represent these classes, we associate with each partition $\vV_i$ a unique \emph{representative} symbol $e_i$. Two variables $x, y \in \vV_i$ satisfy $x \approx y$ and they refer to the same representative $e_i$.

\begin{examplex}\label{ex:union_find}
Consider two union-find data structures over variable set $\vV =\{s,t,x,y,z\}$:
\begin{align*}
 &  \mathbf{a}: \{ e_1: \{x, y, z\}, e_2: \{s, t\}\} & \mathbf{b}: \{e_1: \{x, y\}\}
\end{align*}
$\mathbf{a}$ represents equivalent relations $x\approx y\approx z \land s\approx t$. $\mathbf{b}$ only expresses $x \approx y$.
\end{examplex}

\subsection{union-find data-structure} The basic operations of a union-find data structure are: (1) $\mathit{make}(v)$, which constructs a new equivalence class for a fresh variable $v$; (2) $e := \mathit{find}(v)$ that finds the equivalence class $e$ of a variable $v$; (3) $e := \mathit{union}(x, y)$ that merges equivalence classes containing $x$ and $y$. We additionally provide a special operation (4) $s := \mathit{vars}(e)$, which returns a set $s$ including the variable members within the equivalence classes $e$.

\subsection{Domain Operators} In terms of union-find data structure, our abstract domain operators are defined as follows:
\paragraph{Partial Order} follows the property: a union-find object $\mathbf{m}$ is a \emph{refined} partition of $\mathbf{n}$ if and only if $\forall x, y \in \mathcal V.\; x \approx y$ entailed by $\mathbf{n}$ is also satisfied in $\mathbf{m}$. Thus, $\mathbf{m} \sqsubseteq^\# \mathbf{n} \Longleftrightarrow \left( \forall x, y \in \mathcal V \cdot \; \mathbf{n}.\mathit{find}(x) = \mathbf{n}.\mathit{find}(y) \implies \mathbf{m}.\mathit{find}(x) = \mathbf{m}.\mathit{find}(y)\right)$. As seen in~\cref{ex:union_find}, $\mathbf{a} \sqsubseteq^\# \mathbf{b}$.

\paragraph{Join} is similar to the join of congruence closures algorithm described in~\cite{DBLP:conf/fsttcs/GulwaniTN04}. The result of join regarding union-find is the \emph{least} common partition of both $\mathbf{m}, \mathbf{n}$. That is, $\forall x, y \in \mathcal V.\; x \approx y$ from the result union-find is satisfied in both. The join algorithm, as presented in~\cref{algo:eq_joinmeet}, computes the intersection of two variable sets corresponding to equivalence classes in $\mathbf{m}, \mathbf{n}$ if these two classes have common variables. The result will be further added into $\mathbf{r}$ as a new equivalence class.

\begin{figure}[t]
\begin{algorithmic}
\Procedure{join}{$m, n$}
    \State $r := \{\}$
    \ForAll{$x, y \in \mathcal V_{m} \cup \mathcal V_{n}$}
        \If{$m.\mathit{find}(x) = m.\mathit{find}(y) \land n.\mathit{find}(x) = n.\mathit{find}(y)$}
            \State $e_{\mathbf{m}} := m.\mathit{find}(x)$; $e_{\mathbf{n}} := n.\mathit{find}(x)$
            \State $S_{\mathbf{m}} := m.\mathit{vars}(e_{\mathbf{m}})$; $S_{\mathbf{n}} := n.\mathit{vars}(e_{\mathbf{n}})$
            \State $S_{\mathbf{r}} := S_{\mathbf{m}} \cap S_{\mathbf{n}}$
            \If{$S_{\mathbf{r}} \neq \emptyset$}
                \State pick an element $v\in S_{\mathbf{r}}$
                \State $r.\mathit{make}(v)$
                \ForAll{$v'\in S_{\mathbf{r}} \land v'\neq v$}
                    \State $r.\mathit{make}(v')$; $r.\mathit{union}(v, v')$
                \EndFor
            \EndIf
        \EndIf
    \EndFor
    \State $\Return ~ r$
\EndProcedure

\State
\Procedure{meet}{$m, n$}
    \State $r := m$ \Comment{r is copied from m}
    \ForAll{$x, y \in \mathcal V_{m} \cup \mathcal V_{n}$}
        \If{$n.\mathit{find}(x) = n.\mathit{find}(y)$}
            $r.\mathit{union}(x, y)$
        \EndIf
    \EndFor
    \State $\Return ~ r$
\EndProcedure
\end{algorithmic}
\caption{The join and meet operations.}
\label{algo:eq_joinmeet}
\end{figure}

\paragraph{Meet} computes the \emph{greatest} common partition of $\mathbf{m}$ and $\mathbf{n}$, ensuring that $\forall x, y \in \mathcal V.\; x \approx y$ persisted in the result must also be valid in either $\mathbf{m}$ or $\mathbf{n}$. The meet algorithm is outlined in~\cref{algo:eq_joinmeet} and involves: (1) duplicating the union-find object $\mathbf{m}$ to serve as $\mathbf{r}$; (2) iterating over each pair of equivalence relation $x\approx y$ in each equivalence class $e_{\mathbf{n}}$ of $\mathbf{n}$ to further refine the classes in $\mathbf{r}$. Refer back to~\cref{ex:union_find}, $\mathbf{a} \sqcap^\# \mathbf{b} = \mathbf{a}$.

\neweqdomain has no infinite increasing (decreasing) chains, so the widening (narrowing) operation follows join (meet) strictly. Next, we provide auxiliary operations related to the transfer function in~\cref{sec:abs-sem}.

\begin{figure}[t]
\begin{algorithmic}
\Procedure{$\addEquality$}{$m, x, y$}
    \State $r := m$ \Comment{r is copied from m}
    \State $r.\mathit{make}(y)$
    \State $r.\mathit{union}(x, y)$
    \State $\Return ~ r$
\EndProcedure
\State
\Procedure{$\testEquals$}{$m, x, y$}
    \State $\Return ~ m.\mathit{find}(x) = m.\mathit{find}(y)$
\EndProcedure
\end{algorithmic}
\caption{The $\addEquality$ and $\testEquals$ operations.}
\label{algo:eq_auxiliary}
\end{figure}

\paragraph{$\addEquality$} adds a fresh variable $y$ (which does not belong to any equivalent class) into a class of another variable $x$ in value $m$. The procedure is shown in \cref{algo:eq_auxiliary}.

\paragraph{$\testEquals$} checks whether two variables $x$ and $y$ are equals or not in value $m$, shown in \cref{algo:eq_auxiliary}. Regarding union-find data structure, we test equality by checking whether or not two variables are in the same class. 

\paragraph{$\m{toCons}$} coverts a union-find data structure $m$ into a conjunction of equalities over variables. The function iterates each equivalence class and pairs each variable with every other variable in the class as one equality (known as a linear constraint). As shown in \cref{ex:union_find}, $\m{toCons}(\mathbf{a})$ will compute a system of equality constraints $\{x = y, y = z, x = z, s=t\}$. 

\end{document}